\documentclass[journal]{IEEEtran}
\usepackage{amsmath}
\usepackage{bm}
\usepackage{array}
\usepackage[caption=false,font=footnotesize]{subfig}
\usepackage{cite}
\usepackage{graphicx}
\usepackage{url}
\usepackage{mathrsfs,amsmath}
\usepackage{float}
\usepackage[usenames, dvipsnames]{color}
\usepackage{algorithmic}
\usepackage[linesnumbered, ruled, norelsize, algoruled, boxed, lined]{algorithm2e}
\usepackage{multirow}
\usepackage{tabularx,ragged2e,booktabs}

\SetKwRepeat{Do}{do}{while}%
\newcommand{\removelatexerror}{\let\@latex@error\@gobble}
\newcommand{\nosemic}{\renewcommand{\@endalgocfline}{\relax}}
\newcommand{\dosemic}{\renewcommand{\@endalgocfline}{\algocf@endline}}
\let\oldnl\nl
\newcommand{\nonl}{\renewcommand{\nl}{\let\nl\oldnl}}
\makeatother

\begin{document}

\title{R-FFT: Function Split at IFFT/FFT in Unified LTE CRAN and
  Cable Access Network\thanks{Please direct correspondence to
    M.~Reisslein.}}

\author{Akhilesh~S.~Thyagaturu, Ziyad Alharbi, and~Martin~Reisslein
\thanks{A.~Thyagaturu, Z.Alharbi, and M.~Reisslein
are with the School of Electrical,
Computer, and  Energy Eng., Arizona State Univ., Tempe, AZ 85287-5706, USA
(e-mail: \{athyagat, zalharbi, reisslein\}@asu.edu).}
\thanks{Z.~Alharbi is also with the King Abdulaziz City for Science and Technology, Riyadh, Saudi Arabia, (e-mail: zalharbi@kacst.edu.sa).} }

\maketitle

\begin{abstract}
The Remote-PHY (R-PHY) modular cable network for Data over Cable
Service Interface Specification (DOCSIS) service conducts the physical
layer processing for the transmissions over the broadcast cable in a
remote node.  In contrast, the cloud radio access network (CRAN) for
Long-Term Evolution (LTE) cellular wireless services conducts all
baseband physical layer processing in a central baseband unit and the
remaining physical layer processing steps towards radio frequency (RF)
transmission in remote nodes.  Both DOCSIS and LTE are based on
Orthogonal Frequency Division Multiplexing (OFDM) physical layer
processing.  We propose to unify cable and wireless cellular access
networks by utilizing the hybrid fiber-coax (HFC) cable network
infrastructure as fiber fronthaul network for cellular wireless
services.  For efficient operation of such a unified access network,
we propose a novel Remote-FFT (R-FFT) node that conducts the physical
layer processing from the Fast-Fourier Transform (FFT) module towards
the RF transmission, whereby DOCSIS and LTE share a common FFT module.
The frequency domain in-phase and quadrature (I/Q) symbols for both
DOCSIS and LTE are transmitted over the fiber between remote node and
cable headend, where the remaining physical layer processing is
conducted.  We further propose to cache repetitive quadrature
amplitude modulation (QAM) symbols in the R-FFT node to reduce the
fronthaul bitrate requirements and enable statistical multiplexing.
We evaluate the fronthaul bitrate reductions achieved by R-FFT node
caching, the fronthaul transmission bitrates arising from the unified
DOCSIS and LTE service, and illustrate the delay implications of
moving part of the cable R-PHY remote node physical layer processing
to the headend.  Overall, our evaluations indicate that the proposed
R-FFT node can effectively support unified DOCSIS and LTE services
over the HFC cable plant while substantially reducing the fronthaul
bitrate requirements of the existing CRAN structures.
\end{abstract}


\section{Introduction}

\subsection{Motivation: Modular Cable DOCSIS and Cellular LTE Architectures}
The architectures of both the broadcast cable DOCSIS access network
and the cellular wireless LTE access network have recently been
evolving towards modular architectures.  In broadcast cable networks,
the Modular Headend Architecture version 2 (MHAv2)~\cite{MHAv2}
implements the Cable Modem Termination System (CMTS) functions in a
modular fashion.  Specifically, in the R-PHY
architecture~\cite{alh2017per,Chapman2014,Chapman2015}, a digital
fiber links the headend with distributed Remote PHY nodes (RPDs).  An
RPD can be located close to the cable modems (CMs), improving the
signal quality on the broadcast cable.  The RPD conducts all the
physical layer processing for the transmissions to and from the CMs,
while the higher layer processing is conducted centrally at the
headend.

Similarly, in cellular wireless access, the Cloud Radio Access Network
(CRAN) architecture splits communication functions between centralized
Base Band Units (BBUs) that conduct the baseband signal processing and
Remote Radio Units (RRUs), which conduct the passband processing for
the physical RF transmissions.  A central BBU can support multiple
RRUs and thus provide a common platform for centralized resource
management.  BBUs are typically flexibly implemented in software on
generic computing
hardware~\cite{dai2014uplink,galis2013softwarization} and are amenable
to implementation on cloud computing resources.  Also, conducting the
baseband processing in the BBU reduces the complexity and cost of the
RRUs, which is particularly advantageous for large-scale small cell
deployments.

Importantly, both DOCSIS~3.1~\cite{Hamzeh2015} and LTE are based
on OFDM physical layer processing, which requires an
IFFT/FFT module as main last step of the baseband processing.
In the downstream direction, the FFT module produces the time domain
I/Q samples that the LTE CRAN transports over the fronthaul link
from BBU to RRUs.

\subsection{Challenge: Fronthaul for CRAN}
A critical challenge of CRAN operation is the fronthaul transport of the time
domain I/Q samples between BBU and RRUs, which require low latency and
high bitrates~\cite{Wubben2014}.  A low-latency high-bitrate
connection must constantly be maintained between BBU and RRUs,
regardless of the actual user traffic.  That is, the analog RF signals
must be transmitted and received at all times, even when there is no
wireless user activity. For instance, the passband signal with the
cell broadcast information and reference or pilot tones must always be
transmitted.  Thus, the I/Q samples of the RF passband must be
transported at the constant rate at all times.  Moreover, the
transmission requirements over the optical fiber increase linearly
with the number of remote nodes.  Therefore, numerous techniques, such
as~\cite{nieman2013time, nanba2013new, guo2013lte, guo2012cpri}, have
been proposed to dynamically compress the RF I/Q samples for
effective transmissions over the optical fiber.  However, the
compression techniques are lossy because of the RF signal
quantization, reducing the sensitivity of the receiver in the
upstream.  Nevertheless, the data rate requirements between BBU and
RRUs typically lead to dedicated costly deployments of optical fiber
connections and static allocations of transmission resources.

\subsection{Solution: Unify Cable DOCSIS and CRAN LTE Networks}
We address the LTE CRAN fronthaul challenge by exploiting the fiber
capacity between the cable headend and the cable remote nodes in the
installed hybrid fiber-broadcast coax networks.  In particular, we
propose a novel Remote-FFT (R-FFT) architecture, see
Section~\ref{sec:common:infra}, that co-locates the LTE RRUs with the
cable remote nodes, while the LTE BBUs are co-located with the cable
headends (or outsourced to a cloud resource).  The R-FFT node includes
the IFFT/FFT module as well as the conventional RRU processing modules
towards the RF transmission.  Both cable DOCSIS and wireless LTE share
the IFFT/FFT module in the R-FFT node.  Thus, both DOCSIS and LTE
frequency domain I/Q samples are transported over the fiber link
between cable headend and remote nodes.
In order to further reduce the fronthaul bitrates in the downstream
direction, we propose to cache repetitive QAM symbols in the remote
nodes in Section~\ref{sec:caching}.  Our evaluations in
Section~\ref{sec:perf:eval} present the bitrate reductions achieved
by the QAM symbol caching in the remote node. We also evaluate the
fronthaul bitrates required for the unified DOCSIS and LTE operation
and the delay implications of the transition of an existing R-PHY
cable remote node to an R-FFT remote node.

\subsection{Related Work}
Our study relates to modular access network strategies that have so far
mainly been
studied in isolation for broadcast cable networks and
for wireless cellular networks as well as to caching strategies in
access networks.  Broadcast cable access networks have been
extensively studied for providing wired broadband Internet access to
residential users~\cite{bha2005emp,ful2005doc,hey2010sta,kuo2003imp,
  lee2007eff,lia2006beh,mar2007sim,sha2011per,she1999new,wan2003pac}.
Recent studies have examined the impact of the distance between the
remote node and cable headend on the medium access control (MAC)
performance of the R-PHY modular architecture (which conducts all
physical layer processing in the remote node and the MAC in the
headend) and the R-MACPHY architecture (which conducts all physical
layer processing plus the MAC in the remote
node)~\cite{alh2017per,Chapman2015}.  The studies found that the
modular R-PHY architecture gives good throughput-delay performance for
short headend-to-remote node distances up to around 100~km.  We
consider the R-PHY modular architecture as starting point for our
R-FFT node development and move some of the physical layer processing
to the headend.

A very extensive set of literature has examined modular wireless
cellular access network architectures.  Extensive CRAN studies have
demonstrated the advantages and challenges of conducting the LTE
physical layer baseband processing in the
BBU~\cite{chang2016impact,checko2015cloud,fernando2013mobile,liu2016sta,peng2016recent,wu2015cloud,zha2017coa}.
The high transmission bitrate requirements for transporting the
time-domain I/Q samples produced by the baseband processing at the BBU
to the RRU have spurred research on fronthaul transport strategies,
see e.g.,~\cite{miyamoto2016analysis,miy2017per,ton2016min}, and
alternative function splits between BBU and
RRU~\cite{bov2016phy,cha2017fle,dot2013qua,maeder2014towards,mou2017fea,mak2016exp,nik2017tow,nishihara2016study,wan2017fro,wan2017int,Wubben2014}.
Complementary to this extensive research, which has examined cellular
wireless access in an isolated manner and typically considered
abstract models for the fronthaul between BBU and RRU, we propose to
unify cable and wireless access networks.  More specifically, we
pursue a specific function split at the IFFT/FFT module that $(i)$
reduces the fronthaul transmission bitrate requirements compared to
the conventional CRAN time domain I/Q sample transmission, $(ii)$
shares the IFFT/FFT module in the remote node among DOCSIS and LTE,
and $(iii)$ shares the fiber infrastructure between cable headends and
remote nodes for DOCSIS transport and LTE cellular wireless fronthaul.

Mechanisms to reduce the carbon foot print of access networks have
recently been investigated in wireless
networks~\cite{de2014enabling,fiorani2016joint,kan2012ene,kan2012des,
  oh2013dynamic,suarez2012overview,wu2015energy} as well as cable
networks~\cite{zhu2011novel,Zhu2012,Lu2013}.  This line of energy
saving research has included studies on the caching of application
layer content items in or near the RRUs,
e.g.,~\cite{li2017caa,stephen2016green,zha2016clu}.  In contrast to
the caching of application layer content items, we examine the
caching of repetitive PHY layer QAM I/Q symbols at the RRU.

Only few studies have explored supporting wireless services with cable
networks.  In particular, the channel propagation characteristics of
indoor femto cells that are supported over cable links have been modeled in~\cite{Gambini2012,gambini2010lte}.  The economic
benefits of general infrastructure sharing by residential wired and
cellular wireless networks have been explored
in~\cite{pereira2012infrastructure}. The economic benefits of
integrating LTE and DOCSIS have been discussed
in~\cite{bri2014imp,gruber2014broadband}, while general fiber cost
sharing has been studied in~\cite{sch2014cos}.  We note for
completeness that the application layer performance of LTE wireless
access has been compared with wired DOCSIS access
in~\cite{bec2014mea}; however, the study~\cite{bec2014mea} did not
seek to unify LTE and DOCSIS networks.  In contrast to the existing
studies, we seek to efficiently unify DOCSIS cable and LTE wireless
access networks through the sharing of the cable headend-to-remote
node fiber infrastructure and the sharing of the IFFT/FFT module in
the remote node.  At the same time, the PHY layer function split at
the IFFT/FFT module reduces the high fronthaul bitrate requirements of
the conventional CRAN with PHY layer baseband processing at the BBU,
while still allowing for extensive softwarized physical layer
processing at the BBU or headend.

\section{Background on Function Splits in LTE and Cable Networks}
\label{sec:backgroud}
\subsection{Wireless Downstream vs. Upstream Transmissions}
\subsubsection{Upstream}
In the upstream direction, the RRU receives the RF signal transmitted
from the users. This analog passband signal is down-converted to the
baseband and digitized for the transmission to the BBU for baseband
processing. Unlike the cable link in traditional cellular networks and
antenna infrastructures, the CRAN connects the BBU and RRU with a
digital optical fiber.  The cable link in traditional infrastructures
added significant attenuation to the upstream signal, which is
especially harmful due to the low signal levels received from the user
devices. In contrast, the digital fiber does not contribute towards
the attenuation loss as it carries the signal in digital form.
Extreme care is needed at the RRUs for digitizing the uplink signal
from the users as any additional loss should be avoided due to the low
level of the uplink RF signal at the RRU. For example, if the cable
link accounts for $2$~dB of loss and the noise floor is $-120$~dB,
then the received signal at the RRU connected to a BBU over a cable link
must be $\geq-118$~dB for successful detection. The received signal
can be as low as $-120$~dB if the RRU is connected to a BBU through a
digital fronthaul link, thus the digital fronthaul link increases the
dynamic range of the system by 2~dB.  The Single Carrier
Orthogonal Frequency Division Multiplexing (SC-OFDM) uplink modulation format
is used in typical current deployments. However, SC-OFDM requires
a Discrete Fourier Transform (DFT) just
before the FFT computation, so as to spread an I/Q sample
across multiple FFT input channels.
This DFT spreading, while ensuring good noise resilience, is a complex
operation. Also, SC-OFDM has spectral inefficiencies.
Therefore, technology is advancing towards
uplink OFDM systems where the complex preprocessing of the
discrete Fourier transform (DFT) before FFT computations can be eliminated, especially for MIMO
applications~\cite{studer2016quantized, pitarokoilis2016performance}.
Therefore, we focus on symmetrical OFDM systems in both the upstream
and downstream directions in this article.
The processing
of the upstream signals for the detection and extraction of
information from the RF uplink signals can be centrally executed in
the cloud based BBU on generic hardware, such as the general purpose processors.

\subsubsection{Downstream}
In the downstream direction, the BBU sends the information to the RRUs
for the generation of the passband signal to transmit over the
physical antennas. The RRUs can easily set the transmit power level
gain states for RF signals.  In contrast to the upstream direction,
there is no significant difference in terms of power level of the
signal generation or the dynamic range of the systems between cable
and digital fronthaul links.  Similar to the centralized processing of
the upstream in the cloud based BBU, the information is centrally processed on
generic hardware, such as general purpose processors, to generate
the baseband downstream signals.

\subsection{Function Split in LTE}
\begin{figure}[t!] \centering
\includegraphics[width=3.5in]{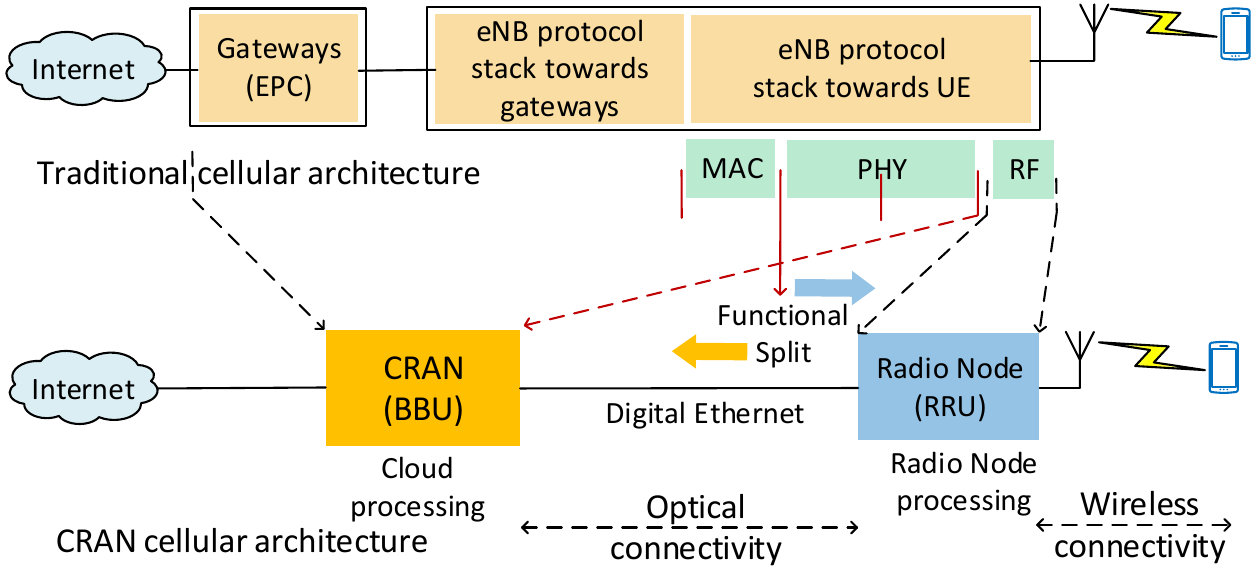}  \vspace{-0.2cm}
\caption{The Cloud RAN (CRAN) implements the RAN functions on the
  cloud-based Baseband Unit (BBU), where baseband signals are
  processed and digital information is transmitted to a remote radio
  node, the Remote Radio Unit (RRU). The RRU generates the passband
  signal for the physical transmission of the wireless RF signal over the
  antenna.  The RAN protocol stack towards the UE, especially at the MAC
  and PHY layers can be flexibly split between BBU and RRU to relax the data
  rate and latency constraints on the optical fiber.}  \label{fig_CRAN}
\end{figure}
Figure~\ref{fig_CRAN} shows the conventional CRAN deployment in
comparison to traditional cellular deployments.  A radio base station
protocol stack, e.g., the LTE protocol stack at the eNB towards the
UE, can be functionally split and implemented flexibly over radio
remote node and BBU.  The conventional CRAN transports the baseband
time domain I/Q samples over optical fiber to the RRUs. The number of
supported RRUs is limited by the amount of traffic over the optical
fiber.  Let $R_o$ [bit/s] denote the capacity of the fronthaul optical
connectivity and $R_u$ denote the data rate required by RRU $u$.
Then, the maximum number of RRUs $N$ that can be supported over the
fronthaul link is the largest $N$ such at $\sum_{u = 1}^N R_{u} \leq
R_o$. In present CRAN deployments, the fronthaul link resources are
typically statically allocated.  Therefore in symmetrical and
homogeneous deployments with equal RRU data rates, i.e., $R_1 = R_2 =
\cdots = R_N = R_{\bar{u}}$, the fronthaul link can support at most $N
= R_o/R_{\bar{u}}$ RRUs. The main bottleneck for CRAN deployments is
the delay and capacity of the fronthaul link.

\begin{table}[t]
\caption{Summary of main notations}
\vspace{-0.2cm}  \label{tab:CRANparams}  \centering
  \begin{tabular}{|l|l|} \hline
\multicolumn{2}{|c|}{\textbf{\rule{0pt}{1\normalbaselineskip}
		Transmission bitrates}} \\
$R_C$  & Cable DOCSIS transmission bitrate (capacity) [bps] \\
$\rho_C$  & Cable payload traffic intensity (load) [unit-free], \\
  & \ \ \ corresp. bitrate is $\rho_C \cdot R_C$ \\
$R_L$  & Wireless LTE transmission bitrate (capacity) [bps] \\
$\rho_L$  & LTE payload traffic intensity (load) [unit-free], \\
  & \ \ \ corresp. bitrate is $\rho_L \cdot R_L$ \\
$R^P$ & Passband bitrate [bps]\\
$R^B$ & Baseband time domain I/Q bitrate [bps] \\
$R^F$ & Baseband frequency domain I/Q bitrate [bps]\\

$R_o$   & Transm. bitrate (capacity) of opt. fiber between headend \\
 & \ \ \ and remote node \\
$\rho_B$   & Traffic intensity from baseband time domain
I/Q samples, \\ & \ \ \ relative to $R_o$ \\  \hline

\multicolumn{2}{|c|}{\textbf{\rule{0pt}{1\normalbaselineskip}
LTE CRAN parameters with typical settings}} \\
$N$ & Number of RRUs per CRAN BBU, $N = 1$ \\
$B_{\rm sub}$  & Number of subcarriers, $B_{\rm sub} = 1200$ \\
		$K$ & Bits per $\Re$/$\Im$ of I/Q sample, $K = 10$ bits   \\
		$W$ & Number of Tx/Rx antennas, $W = 1$ \\
		$T_s$ &  OFDM symbol duration, $T_s = 66.6$ $\mu$s \\
		$f_s$ &  Sampling frequency, $f_s = 30.72$ MHz \\
		$f_c$ &  Carrier frequency, $f_c = 2$ GHz \\ \hline
	\end{tabular}  \vspace{-0.3cm}
\end{table}
To understand the RRU fronthaul requirements, we estimate the data
rates required by the conventional CRAN, where the baseband I/Q
samples are transported from the BBU to the RRU, which is the most
common LTE deployment scenario. General data
rate comparisons of various function split approaches in the LTE
protocol stack have been conducted
in~\cite{dot2013qua,maeder2014towards, Wubben2014}.  Complementary to
these existing evaluations, we closely examine the data rate
requirements based on the implementation specifics of the protocol
stack. That is, we track the information flows across multiple LTE
protocol stack layers and identify the key characteristics that govern
the fronthaul link requirements.  Based on the computationally
intensive FFT operation, the data flow between BBU and RRU can be
categorized into two types: $1)$ time domain samples, and $2)$
frequency domain samples.
Table~\ref{tab:CRANparams} summarizes
the main parameters for the evaluation of the fronthaul optical link
requirements connecting RRU and BBU in the LTE context.
We consider in the following evaluations an LTE system with
20~MHz system bandwidth, which has an $f_s = 30.72$ MHz sampling
frequency and can support an LTE transmission bit rate of
$R_L = 70$~Mbps.

\subsubsection{Time Domain I/Q Sample Forwarding}  \label{sec:time:iq}
The time domain I/Q samples represent the RF signal in the digital
form either in the passband or the baseband. The digital
representation of the passband signal requires a very high data rate
that depends on the physical transmission frequency band. Thus,
passband time domain I/Q sample forwarding is usually
non-economical. For example, in an LTE system, the passband signal is
sampled at twice the carrier frequency $f_c$, with each sample
requiring $K = 10$ bits for digital representation.  Although the LTE
deployment norm is to use $W = 2$ or more eNB antennas, for clarity
and simplified comparison of multiple function split mechanisms, we
set the number of antennas to $W=1$.  The resulting passband I/Q data
rate over the fronthaul link is
\begin{eqnarray}  \label{eq:passband}
R^P &=& N \times W  \times 2 \cdot f_c \times K   \nonumber \\
  &=& 1 \times 1 \times 2 \cdot 2 \cdot 10^9~\mbox{Hz} \times 10~\mbox{bit} = 40~\text{Gbps}.
\end{eqnarray}

The baseband signal for an OFDM symbol in the time-domain consists of
a number of time samples equal to the number of OFDM subcarriers
because of the symmetric input and output samples of the IFFT/FFT
structure. A cyclic prefix is added to the OFDM signal to avoid
inter-symbol interference. In order to reduce the constraints on the
RF signal generation at the RRU, the baseband signal is sampled at a
frequency of $f_s = 30.72$~MHz, with each sample requiring $K = 10$ bits
for digital
representation, and an oversampling factor of 2.  The resulting
baseband I/Q data rate is
\begin{eqnarray} \label{eq:baseband}
R^B &=& N \times W  \times 2 \cdot f_s \times 2 \cdot K \nonumber \\
&=& 1 \times 1 \times 2 \cdot 30.72 \cdot 10^6~\mbox{Hz} \times 2 \cdot 10 \mbox{bit}
       \label{RBi:eqn}  \nonumber \\
 &=& 1.23~\text{Gbps}. \end{eqnarray}
Although the baseband I/Q data rate $R^B$ is significantly lower than the passband I/Q rate $R^P$, the baseband I/Q data
rate $R^B$ scales linearly with the number of antennas and the
bandwidth. Thus, for large numbers of antennas $W$ and wide aggregated
bandwidth, the baseband data rate $R^B$ can be very high.

\subsubsection{Frequency Domain I/Q Sample Forwarding}  \label{sec:freq:iq}
In a 20 MHz LTE system, the duration $T_s$ of one OFDM symbol,
including the cyclic prefix, is $71.4~\mu$s, which corresponds to 2192
time samples for each $T_s$.  The useful symbol duration in the OFDM
symbol duration $T_s$ is $66.7~\mu$s or 2048 samples, out of which the
cyclic prefix duration is $4.7~\mu$s or 144 samples.  Thus, each set
of 2048 samples in an OFDM symbol (excluding the cyclic prefix)
corresponds to $B_{\rm sub} = 2048$ subcarriers  when transformed by
the FFT.  However, only 1200 of these subcarriers are used for signal
transmission, which corresponds to 100 resource blocks (RBs) of 12
subcarriers; the remaining subcarriers are zero-padded and serve as
guard carriers. This leads to $(2048-1200)/2048=0.41=41~\%$ of unused
guard carriers.  Each OFDM subcarrier is modulated by a complex value
mapped from a QAM alphabet. The LTE QAM alphabet size is based on QAM
bits, such as 64 QAM and 256 QAM.
The resulting frequency domain subcarrier information data rate $R^F$
is proportional to the number of
subcarriers $B_{\rm sub}$.  That is, a vector of complex valued QAM
alphabet symbols of size $B_{\rm sub}$ needs to be sent once every
OFDM symbol duration $T_s$, resulting in the data rate
\begin{eqnarray} \label{eq:freq:symb}
R^F &=& N \times W \times B_{\rm sub} \times T_s^{-1} \times 2 \cdot K
\nonumber \\
&=& 1 \times 1 \times 1200 \times (66.7\cdot 10^{-6}~\mbox{s})^{-1} \times 2 \cdot10~\mbox{bit}
   \nonumber \\
&=& 360~\text{Mbps},  \label{RFi:eqn}
\end{eqnarray}
which is a 70~\% reduction compared to the time domain baseband I/Q
data rate $R^B$.

\subsection{Function Split in Cable Distributed Converged Cable Access Platform (DCCAP) Architectures}
\begin{figure}[t!] \centering
\includegraphics[width=3.3in]{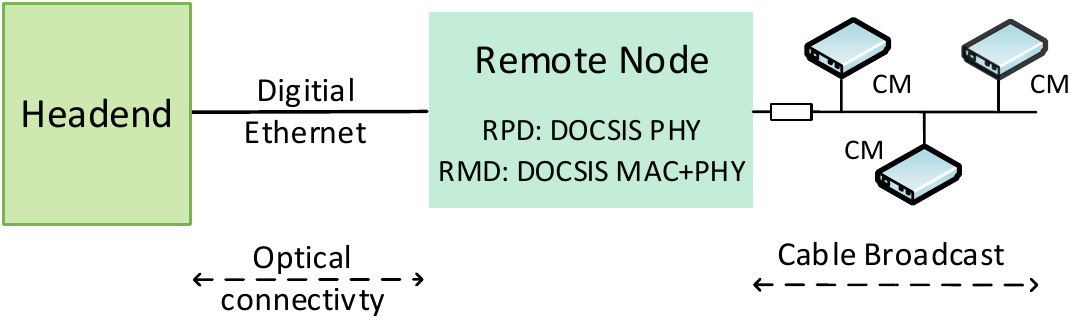}
\vspace{-0.2cm}
\caption{The Distributed Converged Cable Access Platform (DCCAP)
  Architecture separates the modular CMTS functions and implements
  some functions at remote nodes deployed close to the users.  A
  remote node can function as a Remote-PHY Device (RPD) implementing
  the CMTS DOCSIS PHY, or as a Remote-MACPHY Device (RMD) implementing
  the DOCSIS MAC and PHY.}  \label{fig_remote_node}
\end{figure}
The traditional HFC network CCAP architecture implements the CMTS at
the headend and transports the analog optical signal to a
remote node over the optical fiber.  The remote node then converts the
optical analog signal to an electrical RF signal for transmission over
the broadcast cable segment. However, the analog signal is prone to attenuation
in both the optical fiber segment as well as the cable segment.  If
the remote node is deployed far from the headend, then the attenuation of
the optical signal will dominate; conversely, if the remote node is deployed
far from the CMs (users), then the attenuation of the RF signal in the
cable will dominate.

The Modular Headend Architecture (MHA) overcomes the analog optical
signal attenuation in the CCAP architecture by
splitting the CMTS functions, i.e., by modularizing the
implementation of the CMTS functions. The implementation of modular
CMTS functions in a distributed manner across multiple nodes results
in distributed DCCAP architectures~\cite{DCA,Sundaresan2015}.
As shown in Fig.~\ref{fig_remote_node}, the DCCAP architecture defines a
remote node that is connected to the headend through a
digital Ethernet fiber. The digital connection between the remote node and the
headend eliminates the optical signal attenuation, allowing the remote node to
be deployed deep into the HFC network. The remote node deployment deep into the
HFC network reduces the cable segment length, which in turn reduces
the analog RF signal attenuation and improves the overall Signal to
Noise Ratio (SNR) at the CM.  The network connecting the remote node to the
headend is referred to as Converged Interconnect Network (CIN).  The
MHA version 2 (MHAv2)~\cite{MHAv2} architecture defines two DCCAP
architectures: Remote-PHY and Remote-MACPHY.

In the R-PHY architecture~\cite{MHAv2}, the DOCSIS PHY functions in
the CMTS protocol stack are implemented at the remote node, which is
referred to as Remote-PHY Device (RPD). All higher layers in the CMTS
protocol stack, including the MAC as well as the upstream scheduler,
are implemented at the headend.  A virtual-MAC (vMAC) entity can
virtualize the DOCSIS MAC on generic hardware, which can be flexibly
deployed at either the headend or in a cloud/remote data center.  The
RPD is simple to implement and hence has low cost.

\section{Proposed Unified Access Network Architecture for
  LTE and Cable Networks}  \label{sec:common:infra}
\begin{figure}[t!] \centering
\includegraphics[width=3.5in]{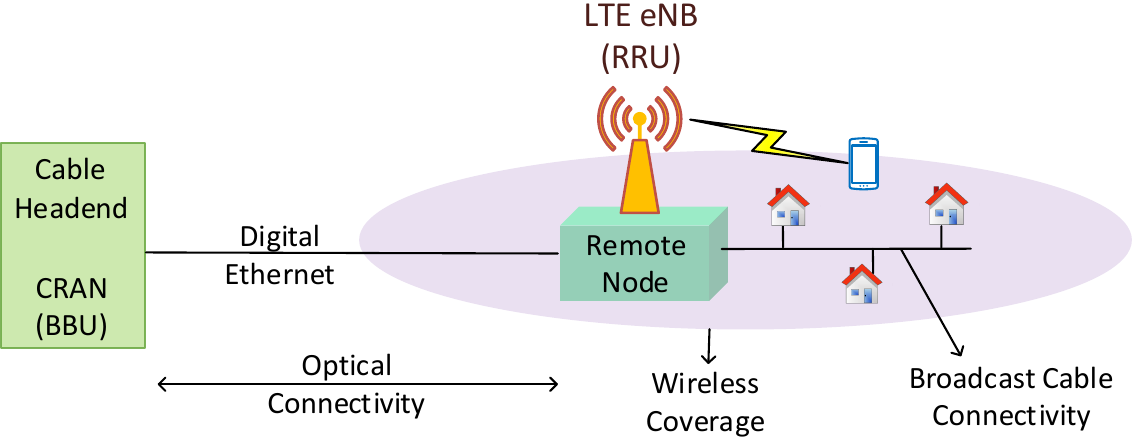} \vspace{-0.2cm}
\caption{Proposed unified LTE and cable (HFC) access network
  architecture: An LTE eNB RRU is deployed at the HFC remote node
  site.  The RRU is connected with the digital optical fiber segment
  of the HFC network, (which functions now also as the radio access
  fronthaul link) to the Base Band Unit (BBU).  The BBU and the Radio
  Access Network (RAN) functions are implemented at the cable headend
  as a cloud RAN (CRAN).}     \label{fig_LTE_cable}
\end{figure}
The digital optical remote node in the DCCAP architecture is deployed
close to the CMs (users). The close proximity of the remote node to
the residential subscribers can be exploited for establishing wireless
LTE connectivity through deploying an LTE eNB RRU at the remote node
site, as illustrated in Fig.~\ref{fig_LTE_cable}.  With the
establishment of LTE connectivity by the cable system operator, users
can be wirelessly connected to the cable system core network for
Internet connectivity, increasing the cable system service
capabilities.  The LTE eNB RRU at the remote node reuses the existing
HFC infrastructure, enabling cable system operators to provide
additional LTE services with low costs.

\subsection{PHY Function Split at IFFT/FFT}
\begin{figure}[t!] \centering
\includegraphics[width=3.5in]{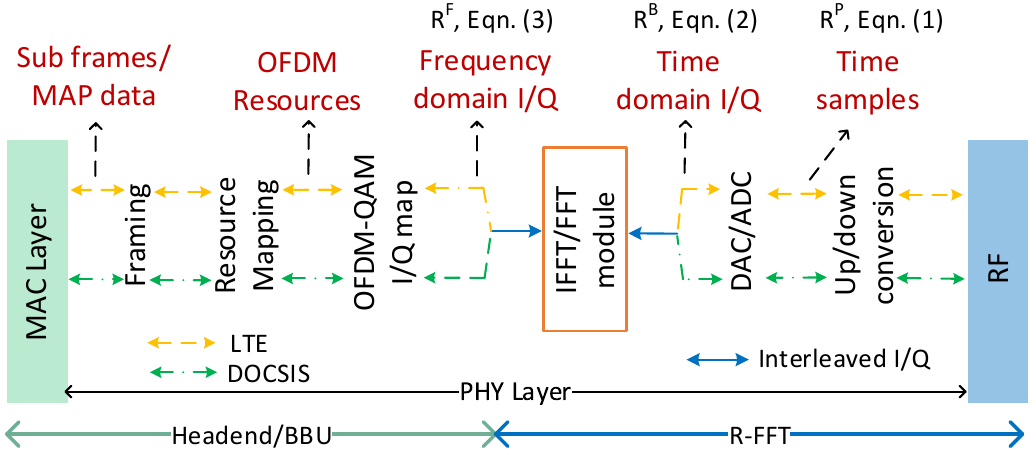}   \vspace{-0.2cm}
\caption{Wireless LTE and cable DOCSIS 3.1 share the same OFDM physical
  layer structure. OFDM based physical layer
  processing can be separated into functions of framing, resource
  mapping, and OFDM-QAM I/Q mapping, which are conducted separately
  for DOCSIS and LTE at the headend in the R-FFT architecture.
  IFFT/FFT processing, Digital to Analog conversion, and passband RF
  signal up/down conversion are then conducted in the R-FFT node.  The
  MAC layer emanates PHY layer payload traffic bitrates $\rho_L R_L$
  for LTE and $\rho_C R_C$ for cable. These payloads are processed in
  the PHY layer, resulting in the increasingly higher bitrates $R^F$,
  $R^B$, and $R^P$.}  \label{fig_split_phy}
\end{figure}
LTE and DOCSIS 3.1 share similar PHY transceiver characteristics for
the OFDM implementation.  We propose to exploit these PHY transceiver
similarities to simultaneously support LTE and DOCSIS over the HFC
network.  The general overview of the physical layer for LTE and
DOCSIS is shown in Fig.~\ref{fig_split_phy}. In the downstream
direction, the data from the MAC layer is processed to form PHY frames
and mapped to OFDM resource locations, which are then converted to
frequency domain QAM I/Q symbols (see Sec.~\ref{sec:freq:iq}) based on
the modulation and coding schemes. The QAM I/Q symbols are then IFFT
transformed to obtain the complex time domain samples.  These time
domain samples (see Sec.~\ref{sec:time:iq}) are then converted to an
analog RF signal for transmission.  In a conventional CRAN, the remote
node conducts the DAC/ADC and the onward processing steps towards the
RF transmission; the conventional CRAN remote node is therefore also
referred to as R-DAC/ADC node.

The I/Q information undergoes different DOCSIS and LTE protocol
specific processing before (to the left of) the IFFT/FFT module as
well as after (to the right of) the IFFT/FFT module.  However, the
same IFFT/FFT module can be used for the I/Q processing of both DOCSIS
and LTE, as illustrated in Fig.~\ref{fig_split_phy}.  Thus, we can
separate (split) the functions at the IFFT/FFT module. That is, the
IFFT/FFT and the processing steps between IFFT/FFT and RF are
implemented at the remote node; whereas the steps towards the MAC
layer are implemented at the headend.  This function split at the
IFFT/FFT node can simultaneously support LTE and DOCSIS over the HFC
network.

\begin{figure*}[t!] \centering
\includegraphics[width=5.5in]{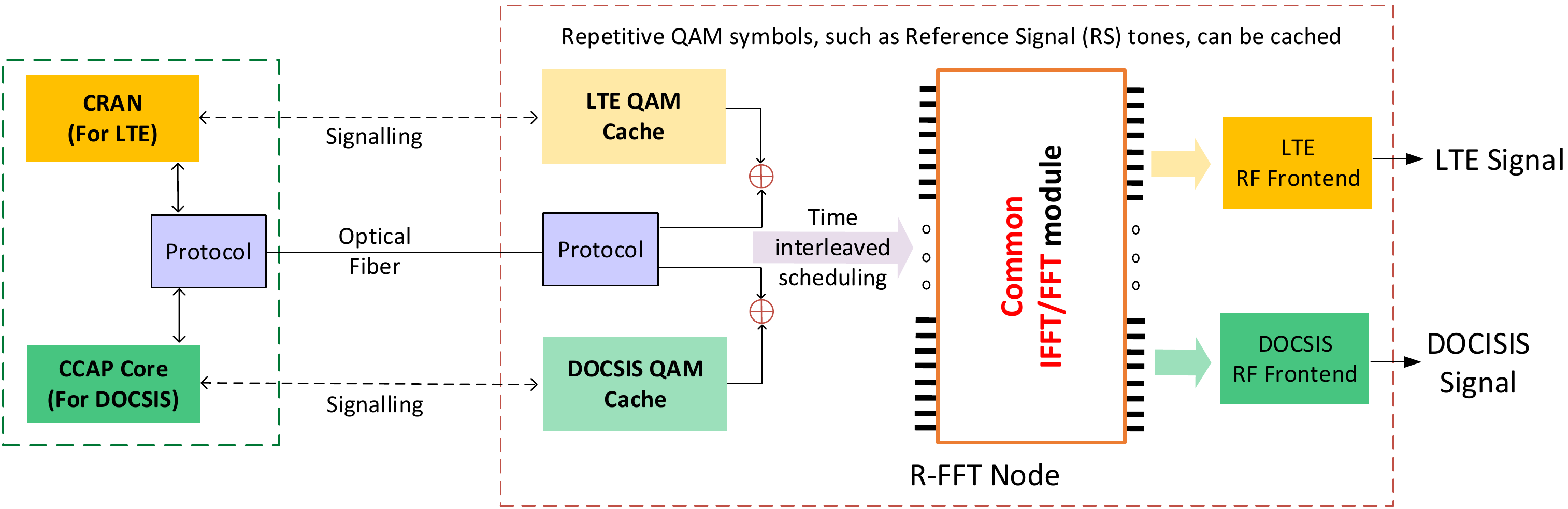}  \vspace{-0.2cm}
\caption{The cable DOCSIS and LTE IFFT/FFT computations are time
  interleaved on a common IFFT/FFT module in the R-FFT
  node. Repetitive DOCSIS and LTE QAM symbols can be cached in the
  R-FFT node, see Section~\ref{sec:caching}, controlled through a
  signalling protocol.}   \label{fig_common_fft}
\end{figure*}
\subsection{Common IFFT/FFT for LTE and DOCSIS}
The LTE and DOCSIS protocols both employ OFDM as the physical layer
modulation technique.  The OFDM modulation relies on FFT
computations~\cite{he1998designing}.  The fact that both LTE and
DOCSIS require the same IFFT/FFT computations for each OFDM modulation
and demodulation can be exploited by using the same computing
infrastructure.  The implementation of parallel FFT computations,
i.e., FFT computations for multiple protocols, on a single computing
infrastructure yields several advantages.  Utilizing the same
computing infrastructure for the LTE and DOCSIS FFT computations
reduces the power consumption and design
space~\cite{khelifi2015parallel,khe2016tow,ma2017analysis,wang2014intel}.

Thus, the main motivation for computing the FFT at the remote node is
to exploit a common remote node platform while flexibly realizing the
different OFDM transmission formats for heterogeneous OFDM based
protocols at the headend.  Figure~\ref{fig_common_fft} illustrates the
R-FFT remote node architecture for simultaneously supporting cable and
LTE.  Generally, in the downstream direction, an IFFT operation is
performed once for every OFDM symbol duration. The LTE OFDM symbol
duration is approximately $T_L = 71.4~\mu$s, while the DOCSIS OFDM
symbol duration is typically either $T_C = 84.13~\mu$s or
40~$\mu$s. However, the actual IFFT compute times $\tau_L$ and
$\tau_C$, for LTE and DOCSIS, respectively, can span from a few
microseconds to several tens of microseconds.  Consequently, there are
typically long idle time periods in the IFFT module inbetween the FFT
computations.  Thus, we can interleave the I/Q input in time such that
same IFFT/FFT module can be used for multiple OFDM based technologies,
e.g., for LTE and DOCSIS.  By reusing the IFFT/FFT computing
structures we can reduce the complexity of the hardware, be more power
efficient, and reduce the cost of the remote node.

\subsection{Proposed Shared Remote-FFT (R-FFT) Node}
In the uplink direction, the proposed R-FFT remote node converts the
incoming DOCSIS RF signal from the CMs to an encapsulated data bits format
that can be transported over the digital fiber link for additional
processing and
onward forwarding at the headend. In a similar way, in the downstream direction,
RF signals are generated from the incoming formatted data bits and
sent out on the RF cable link to the CMs. For LTE, an eNB can use a wide
range of licensed spectrum with a single largest carrier component
of 20~MHz; the bandwidth can be further extended by carrier
aggregation techniques to obtain larger effective bandwidths.
The R-FFT node effectively converts the upstream LTE RF signal
from the wireless users to a digital signal for transport over
the digital fiber link to the
BBU/CRAN. In the downstream direction,
the R-FFT node converts the digital information to an LTE RF
signal for wireless transmission to the users.

We address the high fiber data rate in conventional CRANs through a
balanced split among the functions within the PHY layer while keeping
the remote node simple.  The R-DAC/ADC node in existing conventional
CRANs requires some digital circuitry, such as a CPU, for the DAC and
ADC control.
The FFT/IFFT can be implemented very
efficiently~\cite{jang2016study,kim2017low} so that existing
DAC/ADC remote nodes can take over the FFT/IFFT with relatively
modest modifications or without modifications if the remote node has
enough spare computing capacity.
The advantages of the proposed FFT implementation at the remote node include:
\begin{itemize}
\item[i)] flexible deployment support for LTE and DOCSIS
\item[ii)] requires lower data rate $R^F$, see Eqn.~(\ref{RFi:eqn}), to
  transport frequency domain I/Q samples as compared to time-domain
  I/Q samples, which require the higher $R^B$ rate, see
  Eqn.~(\ref{RBi:eqn}).
\item[iii)] data tones carrying no information are zero valued in the
  frequency I/Q samples, effectively lowering the date-rate over the
  fiber channel for both LTE and DOCSIS, thus enabling statistical
  multiplexing, and
	\item[iv)] possible caching of repetitive frequency QAM I/Q samples,
	such as Reference Signals (RS) and pilot tones.
\end{itemize}

We emphasize that in the proposed R-FFT system, the data rate required
over the fiber is directly proportional to the user traffic. We
believe this is an important characteristic of the FFT function split
whereby we can achieve multiplexing gains by combining multiple R-FFT
nodes, each supporting DOCSIS and LTE services, as illustrated in
Fig.~\ref{fig_LTE_cable}.  In addition, the proposed mechanism enables
the implementation of the complex PHY layer signal processing at the
headend.  Examples of the signal processing operations include channel
estimation, equalization, and signal recovery, which can be
implemented with general-purpose hardware and software. Moreover, the
processing of digital bits, such as for low density parity check
forward error correction, can be implemented at the headend. Thus, the
proposed R-FFT approach reduces the cost of the remote nodes and
increases the flexibility of changing the operational
technologies. The software implementations at the headend can be
easily upgraded while retaining the R-FFT node hardware since the node
hardware consists only of common platform hardware, such as elementary
DAC/ADC and FFT/IFFT components. Thus, the proposed approach eases
technology upgrades. That is, the R-FFT node has minimal impact on
technology advancements because the R-FFT blocks are elementary or
independent of most technology advances.

\subsection{Interleaving Timing of FFT Computations}
\begin{figure}[t!] \centering
\includegraphics[width=3.5in]{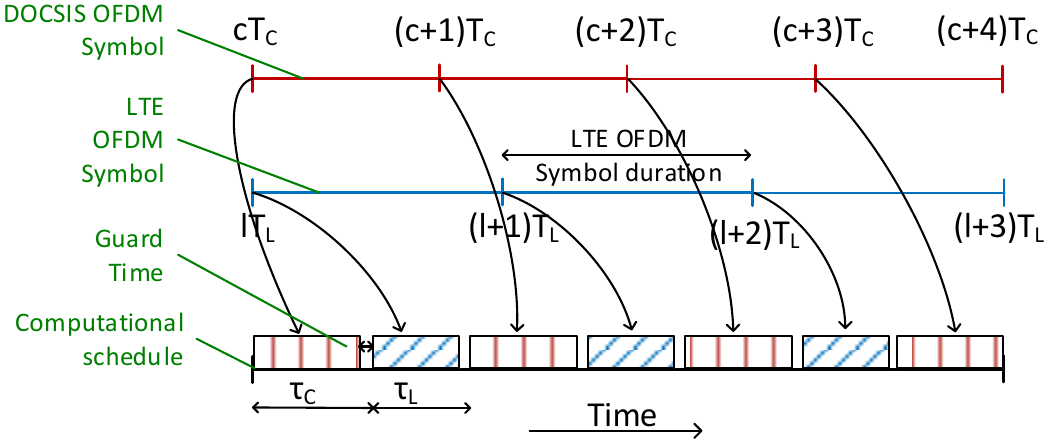} \vspace{-0.4cm}
\caption{The IFFT/FFT computations of two heterogeneous OFDM based
  technologies can be interleaved to use the same computing resource
  at the remote node: Illustration of the periodic cycle behavior
  when long LTE OFDM symbol durations $T_L$ are interleaved with short
  DOCSIS OFDM symbol durations $T_C$, whereby the DOCSIS FFT computation
  takes longer then the LTE FFT computation, i.e., $\tau_C > \tau_L$.}
	\label{fig6}
\end{figure}

In this section we briefly outline the scheduling of the interleaving
of IFFT/FFT computations on a single computing resource.
Sharing a single IFFT/FFT computing module reduces the capital
  and operational expenditures for the remote node compared to
  conducting the DOCSIS and LTE IFFT/FFT computations on two separate
  IFFT/FFT modules.  Nevertheless, we note that it is possible to
  operate an R-FFT node with two separate FFT/IFFT modules.  Such an
  operation with two FFT/IFFT modules would still benefit from the
  lower LTE fronthaul bitrates, but would not achieve the expenditure
  reductions due to sharing a single common IFFT/FFT module.

Figure~\ref{fig6} illustrates the basic timing diagram to
schedule the FFT computations on the computing resource for the case
where $(i)$ the LTE OFDM symbol duration $T_L$ is longer than the
DOCSIS OFDM symbol duration $T_C$ and the DOCSIS FFT computation takes
longer than the LTE FFT computation, i.e., $\tau_C > \tau_L$ (due to
the larger DOCSIS FFT size compared to the LTE FFT).  In
Fig.~\ref{fig6}, $c$ and $l$ denote the indices for the
independent DOCSIS and LTE periodic symbols, which start to arrive
simultaneously at the left edge of the drawn scenario.  We note that
the computation times $\tau_C$ and $\tau_L$ can include a guard time
to account for the context switching between the LTE and DOCSIS
technologies. The switching time depends on the FFT size and
technology-specific parameters, such as the cyclic prefix duration.
The guard time also depends on the memory and CPU functional
capabilities and can vary based on specific implementations.  For
instance, an implementation can include a power savings technique in
which the FFT computing module can be operated in ``sleep modes'',
where, the idle times are power gated to the computing module (i.e.,
the power supply is completely disconnected from the computing
module).  Guard times would then need to compensate for the wake-up
time (i.e., the transition from a sleep mode to a computation mode) in
addition to data load and
read~\cite{fuk20071,fuk20147,hu2004mic,kaw2009sub}.  The resulting
guard times will typically be on the order of microseconds, i.e., a
small fraction of the typical OFDM symbol durations of $40$ and
$80~\mu$secs.

The scheduling of multiple periodic tasks on a shared resource has
been extensively
studied~\cite{but2005rat,how1995non,jos1986fin,shi2008com}.  With
preemptive scheduling, which may interrupt an ongoing computation task,
tasks are schedulable if the sum of the individual ratios of task
computation time to task period duration is less than or equal to
one~\cite{liu1973sch}, i.e., in our context if $\tau_C/ T_C + \tau_L /
T_L \leq 1$.  Non-preemptive scheduling requires an additional
condition~\cite[Theorem 4.1, 2)]{jef1991non}, which in our example
context corresponds to $T_C \geq \tau_L$ in conjunction with $T_L \geq
\tau_L + \tau_C$.  Non-preemptive scheduling appears better suited for
the R-FFT node so as to avoid extra load and read times.
Non-preemptive earliest deadline first scheduling (EDF) can schedule
the tasks that satisfy these preceding conditions.  In particular, we
set the deadline for completing the computation of a symbol arriving
at time $c T_C$, resp., $l T_L$, to be completed by the arrival of the
subsequent symbol at time $(c+1) T_C$, resp., $(l + 1) T_L$.  The
non-preemptive EDF scheduler selects always the tasks with the earliest
completion deadline and breaks ties arbitrarily.  We note that other
schedules could be employed for the relatively simple scheduling of
only two interleaved tasks, e.g., an elementary static cyclic
schedule~\cite{bak1989cyc,sha1989real}.  Additionally, scheduling
techniques that consider energy-efficiency,
e.g.,~\cite{li2013ene,lin2014pra,niu2017ene,yu2016node} may be
considered.  The detailed examination of different scheduling
approaches for the proposed R-FFT node is beyond the scope of this
study and is and interesting direction for future research.
\begin{algorithm}[t!]
	\caption{Caching and FFT Computation Procedure}
	\label{algo:interleaving}
	\SetKwInOut{Input}{input}
	\SetKwInOut{Output}{output}
	\nonl {\bfseries{1. CRAN/Headend}} \newline
	\nl (a) Identify cachable I/Q samples.
          (Secs.~\ref{calte:sec}$+$\ref{cacable:sec}) \newline
	    (b) Create caching rules. (Secs.~\ref{calte:sec}$+$\ref{cacable:sec}) \newline
	    (c) Signal the rules and data for caching. (Sec.~\ref{sec:cache:mgt})	
	\newline
	\nonl \If{Cached I/Q samples require updating}{
		\nonl  Signal remote node for cache renew or flush.
		\nonl }
 \nonl {\bfseries{2. Remote Node}}   \newline
\nonl\ForEach{OFDM Symbol in $T_C$ and $T_L$}{		
  \nonl          \If{Caching is enabled}{
  \nonl				  Read cache and I/Q mapping\;
  \nonl                    Add cache-read I/Q  to received I/Q\;}
  \nonl	    	\If{FFT module is free}{
  \nonl				  Schedule I/Q for FFT\;}
  \nonl          \Else{
  \nonl	Schedule at completion of current execution\;
  \nonl }  \nonl}	   	
\end{algorithm}

The sharing of the FFT/IFFT module by multiple technologies can be
extended to include both upstream and downstream directions, i.e., the
module can be shared by downstream DOCSIS and LTE as well as upstream
DOCSIS and LTE, as the computations for the different directions are
performed independently of each other, even for wireless full-duplex
communications.  Also, the FFT computation duration $\tau$ can
represent the aggregate of multiple OFDM symbol instances.  For
example, in the case of carrier aggregation in LTE (or channel bonding
in DOCSIS), there would be an OFDM symbol for each of the $\alpha$ carrier
component, resulting in $\tau_L=\tau_1 + \tau_2 + \cdots + \tau_{\alpha}$.  Similarly,
computations resulting from multiple LTE eNBs at a single node can be
aggregated and abstracted to a single $\tau_L$.  The proposed approach
can be readily extended to more than two technologies that conduct
their FFT computations by sharing the remote node.

\subsection{Transport Protocols}
A protocol is required to coordinate the I/Q data transmissions over
the transport network. The strict latency requirements for the CRAN
and DCCAP architectures limit the choice of generic protocols over
Ethernet. Some of the fronthaul protocols that could be employed for
the transport of information between headend/cloud and remote node are:

\subsubsection{Radio over Fiber (RoF)}
Radio over fiber (RoF) transports the radio frequency signal over
  an optical fiber link by converting the electrically modulated
  signal to an optical signal~\cite{nir2010dig,nov2016rad,raj2017rev}.
  RoF signals are not converted in frequency but superimposed onto
  optical signals to achieve the benefits of optical transmissions,
  such as reduced sensitivity to noise and interference.  The remote
  nodes directly convert the optical signal to an electrical signal
  with minimal processing, reducing the cost of the remote node.
  However, the analog optical signal transmission in RoF suffers from
  more attenuation as compared to the transmission of digital data
  over the fiber.

We briefly note that so-called Radio-and-Fiber (R\&F) networks are
an alternative form of converged radio and fiber
networks~\cite{ghazisaidi2010fiber,gha2011fib,tsagklas2011survey}.
R\&F networks typically consist of distinct wireless and optical
network segments that each conduct their own specific physical and
medium access control layer processing~\cite{gha2011fib,aur2014fiw,thy2016sof}.
That is, R\&F networks are
typically deployed as two-level architectures with protocol
translation at the interface between the radio and optical network
segments.
For CRAN architectures, the RoF transport is generally preferred over
R\&F networks
as the RoF transport better supports centralized signal processing at the
BBU~\cite{tsagklas2011survey}. The proposed R-FFT approach follows
the generally strategy of the CRAN architecture to centralize signal processing
at the BBU and therefore RoF transport appears
better suited than R\&F networking for the R-FFT approach.

\subsubsection{Common Public Radio Interface (CPRI)}

The eCPRI v1.0 specification~\cite{ecpri,de2016overview,liu2016cpri}
  defines a generic protocol framework for transporting the I/Q
  symbols between the Remote Radio Unit (RRU) and the Base Band Unit
  (BBU) over a conventional transport network, such as Ethernet and
  optical transport networks. The eCPRI framework has been
  fundamentally defined to support a wide range of functional split
  options between RRU and BBU with variable transport bit rates. Our
  proposed R-FFT node corresponds to the low PHY functional split
  option as defined in the eCPRI specification.  The eCPRI protocol
  framework is thus well suited as transport protocol between R-FFT
  node and Headend/BBU.

\subsubsection{Open Base Station Architecture Initiative (OBSAI)}
The Open Base Station Architecture Initiative (OBSAI)~\cite{obsai} is
similar to CPRI in that the digitized time domain I/Q samples are
transported over a fronthaul interface.  The OBSAI would need to be
adapted for the frequency I/Q transport.  In contrast to CPRI, the
OBSAI interface is an IP based connection.  The IP logical connection
can be implemented over any generic Ethernet link, providing flexible
connectivity between headend/cloud and remote node.

\paragraph{External PHY Interfaces}
The Downstream External PHY Interface (DEPI)~\cite{depi} and Upstream
External PHY Interface (UEPI)~\cite{uepi} enable the common transport
mechanisms between an RPD and the CCAP core. DEPI and UEPI are based
on the Layer 2 Tunneling Protocol version 3 (L2TPv3).  The L2TPv3
transparently transports the Layer 2 protocols over a Layer 3 network
by creating pseudowires (logical connections).

We note that the transport between R-FFT remote node and BBU
  needs to also comply with the delay requirements due to the hybrid
  automatic repeat request (HARQ) protocols operating in CRANs. The
  HARQ protocols impose latency requirements that in turn limit the
  distance between RRU and BBU, for
  example~\cite{checko2015cloud,cho2014cost,kim2016mob,nik2015pro}
  consider a 20--40~km RRU-BBU distance.  These delay constraints for
  CRAN networks apply similarly to the R-FFT network.  Fundamentally,
  the R-FFT node only reduces the fronthaul transmission bit rates and
  does not alter the delay requirements of the I/Q transport between
  BBU and RRH.  Therefore, all the CRAN constraints and requirements
  for the delay apply analogously to the R-FFT network.  On the other
  hand, the integration of the cable access network with the CRAN does
  not impose any additional delay requirements (in addition to the
  existing CRAN delay requirements) as there are no HARQ processes in
  the cable network.  Thus, the LTE CRAN requirements dictate the
  delay limits for the combined deployments of CRAN and cable access
  networks in the proposed R-FFT architecture.

\section{Proposed Remote Caching of QAM Symbols}   \label{sec:caching}
\begin{figure}[t!] \centering
\includegraphics[width=3.4in]{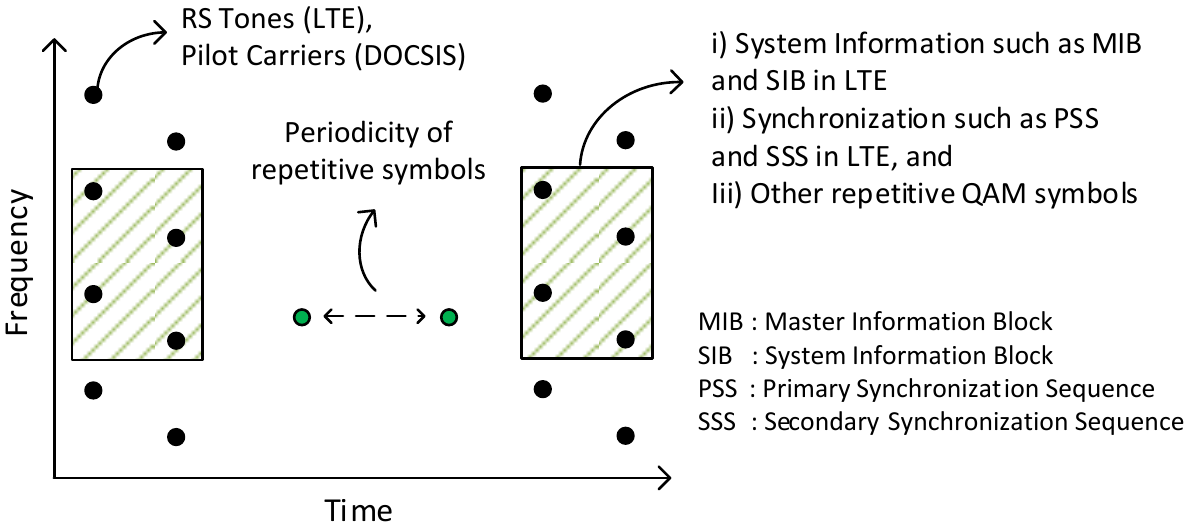}  \vspace{-0.2cm}
\caption{Some QAM symbols pertaining to an OFDM symbol remain constant
  over time.  For example, Reference Signal (RS) tones in LTE occur
  typically every four OFDM symbols; similarly, DOCSIS pilot tones
  repeat periodically.  This periodically repeated QAM symbol
  information can be cached at the remote node to reduce the data rate
  over the digital Ethernet fiber link between the headend and the
  R-FFT node.}  \label{fig_caching}
\end{figure}
In order to further reduce the bandwidth in addition to the function
split process, several techniques, such as I/Q
compression~\cite{nanba2013new, nieman2013time,guo2012cpri,
  joung2013base}, can be employed. In contrast, we propose OFDM
resource element (time and frequency slot) allocation based remote
caching. If some part of the information is regularly and repeatedly
sent over the interface, a higher (orchestration, in case of SDN)
level of the signaling process can coordinate caching mechanisms. For
example, there is no need to transmit the downstream I/Q samples of the
pilot tones as they remain constant in DOCSIS.
Figure~\ref{fig_caching} gives an overview of repetitive QAM symbols
in LTE and DOCSIS.  The stationary resource elements across the time
domain, such as the system information block (SIB), typically change
over long time scales on the order of hours and days. The cached
elements can be refreshed or re-cached through cache management and
signalling protocols, see Sec.~\ref{sec:cache:mgt}.

In contrast to the downstream, upstream information must be entirely
transported to the headend to process all the signal components
received by the R-FFT receiver.

In the evaluations of the overhead due to repetitive QAM symbols that
can be saved through caching in this section, we evaluate the ratios
(percentages) of number of repetitive I/Q symbols to total number of
I/Q symbols.  Subsequently, in the evaluations in
Section~\ref{sec:perf:eval} we evaluate the corresponding reductions
of the fronthaul transmission bitrate.

\subsection{LTE Networks}  \label{calte:sec}
\subsubsection{Reference Signal (RS) Tones Caching}  \label{RStonecache:sec}
RS tones are pilot subcarriers that are embedded throughout the
operational wireless system bandwidth for channel estimation so as to
equalize the impairments of the received wireless signal.
More
  specifically, the RS tones consist of data that is known to both the
  wireless transmitter and receiver, whereby in the downstream
  direction the LTE eNB (RRU) co-located with the R-FFT node is the
  wireless transmitter and the UEs are the wireless receivers. The
  known RS tone data helps the wireless receivers (UEs) to determine
  the downlink power levels (of the signal arriving from the eNB) as well
  as to determine the channel characteristics (distortions)
  by comparing the received RS
  tone signals with the known RS tone data. Importantly, for a given
  cell deployment, the same RS tones are always transmitted at a
  constant power level by the eNB to the UEs so as to facilitate the
  estimation of the received power level and distortion after the
  signal propagation over the wireless channel between eNB and
  UEs. Thus, at the wireless transmitter (eNB) side, the transmitted RS
  tones are constant for a given cell deployment.  Signal
disturbances on the wireless channels are substantially more
pronounced compared to signal propagation in wired
channels. Therefore RS tones are added in close proximity with each
other in LTE to accurately estimate the channel characteristics, such
as coherence-time and coherence-bandwidth. The values and
  positions of the RS tones are fixed for a given deployment, i.e.,
  the RS tones and the corresponding I/Q sample values do not change
  over time for a given wireless cell deployment. In particular, in
  LTE, as defined in the 3GPP specification TS
  36.211~\cite{3gppts36211}, the configuration of the cell-specific RS
  tones depends on the cell identity (ID) of the deployment, which is
  an integer value between 0 and 503.  The cell ID is constant for a
  given physical deployment; hence, the RS tone configuration is also
  constant.  Therefore, the RS tone caching
  at the R-FFT node has to be performed only once
  during the initialization and there is no need for updating the
  cached RS (pilot) tone signals.

For a single antenna, the RS tones are typically spaced six
subcarriers apart in frequency such that eight RS tones exist in a
single subframe (which consists of 14 OFDM symbols in the time
dimension) and a single Resource Block (RB) (which consists of 12 LTE
subcarriers in the frequency dimension).  Thus, with a full RB
allocation, i.e., for a relative payload
data traffic load (intensity) of
$\rho_L = 1$, approximately $8 / (12 \times 14) = 4.7$~\% of I/Q
transmissions over the digital fiber can be saved by caching RS tones
at the remote node, regardless of the system bandwidth.
In general, for a traffic intensity $\rho_L,\ \rho_L \leq 1$,
the overhead due to RS tones in the LTE resource grid is
\begin{equation}
\text{RS Overhead} = \frac{8}{ \rho_L \times 12 \times 14  } = \frac{4.7}{\rho_L}~\%.
\label{eq:rs:rho}
\end{equation}
When the user data traffic is very low, e.g., $\rho_L=0.1$, the
overhead is almost $47~\%$, and similarly when $\rho_L=0.01$ the
overhead becomes $470~\%$.

\subsubsection{PHY Broadcast Channel (PBCH) Caching}
The PHY Broadcast Channel (PBCH) carries the Master Information Block
(MIB) which is broadcast continuously by the eNB regardless of the
user connectivity.  The MIB includes basic information about the LTE
system, such as the system bandwidth and control information specific
to the LTE channel.  The PBCH/MIB always uses the six central RBs (i.e.,
72~subcarriers) for the duration of 4~OFDM symbols to broadcast the
MIB data.  The PBCH space in the resource grid is inclusive of the RS
tones used in the calculation of Eqn.~(\ref{eq:rs:rho}); therefore, the RS
tones need to be subtracted when calculating the MIB overhead.
The PBCH/MIB occurs once every 40~ms and there exist four redundant MIB
  versions.  Once all the four versions are cached, the
I/Q samples corresponding to the MIB PDU remain constant for the deployment and
 no further updates are required. 
  The four redundant MIB versions are broadcast with an offset of 10~ms. Thus, an
PBCH/MIB occurs effectively once in every 10~ms (radio frame). The
PBCH/MIB overhead for an entire 20~MHz system LTE system with
1200 subcarriers, 14 OFDM symbols, and 10 subframes is thus
\begin{equation}
\text{PBCH Overhead} = \frac{6 \times 12 \times 4 - (8 \times 6)}{\rho_L \times
  1200 \times 14 \times 10} = \frac{0.142}{\rho_L}~\%.  \label{eq:mib}
\end{equation}
Alternatively, for a 1.4~MHz system with 72 subcarriers (the lowest
currently standardized LTE bandwidth, which would be used for IoT type
of applications), the overhead increases to
\begin{equation}
  \text{PBCH Overhead}_{1.4\text{MHz}} = \frac{6\times 12 \times 4-(8 \times 6)}
       {\rho_L \times 72 \times 14 \times 10} = \frac{2.3}{\rho_L}~\%. \label{eq:mib:14}
\end{equation}
Future IoT related standardization efforts may lower the LTE rates below
1.4~MHz to better suit the needs of low-rate IoT applications, leading to
further increases of the PBCH overhead.

\subsubsection{Synchronization Channel Caching}
The Synchronization Channel (SCH) consists of the Primary
Synchronization Sequence (PSS) and the Secondary Synchronization
Sequence (SSS), which are broadcast continuously by the eNB,
regardless of the user connectivity. The PSS and SSS help with the
cell synchronization of wireless users by identifying the physical
cell ID and the frame boundaries of the LTE resource grid.
Similar to the RS tones in Section~\ref{RStonecache:sec},
  the PSS and the SSS, i.e., the cell ID and frame boundary information,
  are static for a given cell deployment.
Thus, caching the PSS and SSS does not
degrade the functioning of the LTE cell.
The PSS/SSS occurs every 5~ms (twice per radio frame) and uses six
central RBs over two OFDM symbols. Similar to
Eqns.~(\ref{eq:mib})~and~(\ref{eq:mib:14}), the overhead due to the
PSS/SSS in 20~MHz and 1.4~MHz systems are
\begin{eqnarray}
\text{SCH Overhead} &=& \frac{6 \times 12 \times 4}{\rho_L \times 1200 \times 14 \times 10} = \frac{0.171}{\rho_L}~\%. \nonumber \\
\text{SCH Overh.}_{1.4\text{MHz}}
&=& \frac{6 \times 12 \times 4}{\rho_L \times 72 \times 14 \times 10} = \frac{2.8}{\rho_L}~\%.
\label{eq:sch}
\end{eqnarray}

\subsubsection{System Information Block (SIB) Caching}
In a similar way, the caching mechanism can be extended to the System
Information Blocks (SIBs) broadcast messages of the LTE PHY Downlink
Shared Channel (PDSCH).  There are 13 different SIB types, ranging
from SIB1 to SIB13.
SIB1 and SIB2 are mandatory broadcast messages
that are mostly static for a given cell deployment.  More
specifically, SIB1 contains a System Information info-tag bit. This
info-tag bit changes when the deployment characteristics change, e.g.,
when a new neighbor cell is added or a new feature is added to the
existing cell.  Such changes typically occur only every few weeks or
months. When such a change happens, then the info-tag bit signals that
all SIBs need to be updated.
Similarly, the other SIBs depend on the relations between the serving
  cell
and the neighbor cell configurations. In a typical deployment, SIB3 to
SIB9 are manually configured
and can be combined in a single message block
for the resource block allocation.
Typical RB allocation configurations schedule the SIB1 and SIB2 transmissions over 8 RBs across
14~OFDM symbols in time (i.e., 1~subframe).
with an effective periodicity (with redundant version transmissions)
of 2 radio frames (i.e., 20~ms). The overhead from the
SIB1 and SIB2 transmissions while subtracting the corresponding
RS tones overhead of $8 \times 8$, i.e., 8 tones per RB for 8 RBs, is
\begin{eqnarray}
  \text{SIB Overhead} &=& \frac{8 \times 12 \times 14 - (8 \times 8)}
       {\rho_L \times 1200 \times 14 \times 20} = \frac{0.381}{\rho_L}~\%. \nonumber \\
\text{SIB Overh.}_{1.4\text{MHz}}
&=& \frac{8 \times 12 \times 14 - (8 \times 8)}
     {\rho_L \times 72 \times 14 \times 20} =
\frac{6.3}{\rho_L}~\%.   \label{eq:sib}
\end{eqnarray}
The resource allocation and periodicity of the higher order SIBs,
  i.e., from SIB3 to SIB9, can vary widely and it is therefore
  difficult to accurately estimate the overhead. We consider therefore
  only the SIB1 and SIB2 caching in our evaluation of the cache
  savings.  However, a signalling and cache management protocol, as
  outlined in Section~\ref{sec:cache:mgt}, can coordinate the caching
  of the higher order SIBs and thus achieve further savings. 

\subsection{Cable Networks}  \label{cacable:sec}
In DOCSIS~3.1, downstream pilot subcarriers are modulated by the CMTS
with a predefined modulation pattern which is known to all CMs to
allow for interoperability.  Two types of pilot patterns are defined
in DOCSIS~3.1 for OFDM time frequency grid allocations: i) continuous,
and ii) scattered.  In the continuous pilot pattern, pilot tones with
a predefined modulation occur at fixed frequencies in every symbol
across time.  In the scattered pilot pattern, the pilot tones are
swept to occur at each frequency locations, but at different symbols
across time.  The scattered pilot pattern has a periodicity of 128
OFDM symbols along the time dimension such that the pattern repeats in
the next cycle.  Scattered pilots assist in the channel estimation.
Typical deployments have 192~MHz operational
bandwidth~\cite{cabledeploy}, corresponding to an FFT size of 8192 with
25~kHz subcarrier spacing. A 192~MHz system has 7680 subcarriers,
including 80 guard band subcarriers, 88 continuous pilot subcarriers,
and 60 scattered pilot subcarriers. Therefore, the overhead due to
guard band and pilot subcarriers, which can be cached at the remote
node, is
\begin{equation}
  \text{Cable Over.} = \frac{80 + 88 + 60}{\rho_C \times 7680}
     = \frac{2.9}{\rho_C}~\%.  \label{eq:cable}  \end{equation}

\subsection{Memory Requirements for Caching}
The caching of frequency domain OFDM I/Q symbols requires
caching memory at the remote node.
Each I/Q symbol that needs to be cached is a complex number
with real and imaginary part.
For the purpose of evaluation, we
follow~\cite{Wubben2014,maeder2014towards} and consider
a 10 bit representation for each part of the complex number, resulting
in a 20 bit memory requirement for each frequency domain QAM symbol.
A 30 bit representation of a frequency domain QAM symbol, as considered
in~\cite{ecpri}, would correspondingly increase the memory requirements.
The caching of LTE RS tones saves 4.7~\% of the fronthaul transmissions
as shows in Eqn.~(\ref{eq:rs:rho}).
Within each RB, 8 RS tones
exist for every 12~subcarriers. A typical
20~MHz system with 1200 subcarriers, has thus $8 \times 100$ RS tones. The
total memory required to cache the RS tones QAM symbol data is
\begin{equation}
\centering
\text{RS Tones Mem.} = (8 \times 100) \times 2 \cdot 10~\text{bits} = 16000~\text{bits}.
\label{eq:rs:mem}
\end{equation}
Similarly, caching of the PBCH, SCH, and SIB data requires
\begin{eqnarray}
\text{PBCH Mem.} &=& (6\times12\times4 - (8 \times 6)) \times 2 \cdot 10~\text{bits} \nonumber \\
  &=& 4800~\text{bits}, \\
\text{SCH Mem.} &=& (6 \times 12 \times 4) \times 2 \cdot 10 = 5760~\text{bits},\;\;\: \\
\text{SIB Mem.} &=& (8 \times 12 \times 14-(8 \times 8))
\times 2 \cdot 10~\text{bits} \nonumber  \\
&=& 5760~\text{bits}.
\label{eq:pbch:mem} \end{eqnarray}
For DOCSIS, the cache memory requirement for the continuous and
scattered pilots is
\begin{equation}
\centering
\text{Pilot Tones Mem.} = (80+88+60) \times 2 \cdot 10~\text{bits} = 4560~\text{bits}.
\label{eq:pilot:mem}
\end{equation}
Thus, based on Eqns.~(\ref{eq:rs:rho})--(\ref{eq:cable}), total
savings of approx. 7~\% to 18~\% can be achieved in the fronthaul
transmissions when the full resource allocation ($\rho = 1$) over the
entire bandwidth is considered in both LTE and DOCSIS.  For lower
allocations, i.e., when there is less user data ($\rho < 1$), the
caching can achieve much more pronounced fronthaul transmission
bitrate reductions.  In the extreme case, when there is no user data,
all the cell specific broadcast data information can be cached at the
remote node and the fronthaul transmissions can be completely
suspended.  The total memory for the caching required at the remote
node based on Eqns.~(\ref{eq:rs:mem})--(\ref{eq:pilot:mem}) is less
than 37~kbits.  The implementation of less than 5~kbytes cache memory
at the remote node appears to be relatively simple and no significant
burden for the existing remote nodes. Therefore, we believe that
fronthaul transmission bitrate reductions of more than 7~\% with
almost negligible implementation burden is a significant benefit.

\subsection{Signalling and Cache Management Protocol}  \label{sec:cache:mgt}
The signalling and cache management protocol involves: i) transporting
the caching information to the remote nodes, ii) updating the cached
information at the remote nodes with new information, and iii)
establishing the rules for reading cached resource elements at the
remote node.  Signalling protocol modules at the headend/cloud and
remote node coordinate with each other through a separate (i.e.,
non-I/Q transport) logical connection between the headend/cloud and
remote node, as summarized in Algorithm~1.  Some of the cached
information may change over time; however, these changes occur
typically at much longer timescales compared to the I/Q transmissions
from the headend to the remote node.  Due to the very long time scale
of cache changes, i.e., very infrequent cache changes, the signalling
overhead which arises from the cache management is typically
negligible.  The reading (retrieval) of the cached content has to be
precisely executed with accurate insertion of the subcarrier
information in the particular time and frequency locations.

\section{Performance Evaluation}   \label{sec:perf:eval}

\subsection{Reduction of Downstream Fronthaul Bitrates due to Caching}
\label{eval_br_ca:sec}
\begin{table*}[t!]  \centering
\caption{Downstream LTE fronthaul bitrates without caching
  [$R^F_{\rho, \textrm{Total}}$, Eqn.~(\ref{eq:IQrate})] and with caching
  [$R^F_{\rho, \textrm{Payload}}$, Eqn.~(\ref{eq:totalIQ})], as well as bitrate
  reductions due to I/Q caching with $7\%$ overhead in FFT-split LTE
  system with QAM size $64$ ($2^{6}$), for different payloads $\rho_L$
  and code rates $0.9$, $0.5$, and $0.7$.}
\label{table:tput:lte}
\begin{tabular}{|l|l|l|l|l|l|l|l|}  \hline
\multicolumn{1}{|c|}{\multirow{3}{*}{\textbf{\begin{tabular}[c]{@{}c@{}}LTE \\ Load $\rho_L$ \end{tabular}}}} & \multicolumn{6}{c|}{\textbf{LTE FFT-Split Fronthaul I/Q Data Rate} (\textit{Gbps})}                                                                                                                                                                             & \multicolumn{1}{c|}{\multirow{3}{*}{\textbf{\% Sav.}}} \\ \cline{2-7}
\multicolumn{1}{|c|}{}                                                                              & \multicolumn{2}{c|}{\textbf{Code Rate$ = 0.9$}}                                    & \multicolumn{2}{c|}{\textbf{Code Rate$ = 0.7$}}                                    & \multicolumn{2}{c|}{\textbf{Code Rate$ = 0.5$}}                                    & \multicolumn{1}{c|}{}                                  \\ \cline{2-7}
\multicolumn{1}{|c|}{}                                                                              & \multicolumn{1}{c|}{\textbf{w/o cach.}} & \multicolumn{1}{c|}{\textbf{w/ cach.}} & \multicolumn{1}{c|}{\textbf{w/o cach.}} & \multicolumn{1}{c|}{\textbf{w/ cach.}} & \multicolumn{1}{c|}{\textbf{w/o cach.}} & \multicolumn{1}{c|}{\textbf{w/ cach.}} & \multicolumn{1}{c|}{}                                  \\ \hline
\multicolumn{1}{|c|}{$0.01$}& \multicolumn{1}{|c|}{$0.296$}  & \multicolumn{1}{|c|}{$0.037$} & \multicolumn{1}{|c|}{$0.380$} & \multicolumn{1}{|c|}{$0.047$} &  \multicolumn{1}{|c|}{$0.533$} & \multicolumn{1}{|c|}{$0.066$} & \multicolumn{1}{|c|}{$87.50$}  \\ \hline
\multicolumn{1}{|c|}{$0.1$} & \multicolumn{1}{|c|}{$0.629$}  & \multicolumn{1}{|c|}{$0.370$} &  \multicolumn{1}{|c|}{$0.809$} & \multicolumn{1}{|c|}{$0.476$} & \multicolumn{1}{|c|}{$1.133$} & \multicolumn{1}{|c|}{$0.666$} & \multicolumn{1}{|c|}{$41.17$}  \\ \hline
\multicolumn{1}{|c|}{$0.2$} & \multicolumn{1}{|c|}{$1.000$}  & \multicolumn{1}{|c|}{$0.740$} & \multicolumn{1}{|c|}{$1.285$} & \multicolumn{1}{|c|}{$0.952$} & \multicolumn{1}{|c|}{$1.800$} & \multicolumn{1}{|c|}{$1.333$} & \multicolumn{1}{|c|}{$25.92$}  \\ \hline
\multicolumn{1}{|c|}{$1$}   & \multicolumn{1}{|c|}{$3.962$}  & \multicolumn{1}{|c|}{$3.333$} & \multicolumn{1}{|c|}{$5.095$} & \multicolumn{1}{|c|}{$4.761$} & \multicolumn{1}{|c|}{$7.133$} & \multicolumn{1}{|c|}{$6.666$} & \multicolumn{1}{|c|}{$6.54 $}  \\ \hline	  \end{tabular}           \end{table*}

\begin{table*}[t!]   \centering
\caption{Downstream cable fronthaul bitrates without caching
  ($R^F_{\rho, \textrm{Total}}$) and with caching ($R^F_{\rho,
    \textrm{Payload}}$), and bitrate reductions due to I/Q caching
  with $3\%$ overhead in FFT-split DOCSIS system with QAM size $4096$
  ($2^{12}$), for different packet payloads $\rho_C$ and code rates
  $0.9$, $0.5$, and $0.7$}
	\label{table:tput:docsis} 	\begin{tabular}{|l|l|l|l|l|l|l|l|}
		\hline
\multicolumn{1}{|c|}{\multirow{3}{*}{\textbf{\begin{tabular}[c]{@{}c@{}}DOCSIS \\ Load $\rho_C$ \end{tabular}}}} & \multicolumn{6}{c|}{\textbf{DOCSIS FFT-Split Fronthaul I/Q Data Rate} (\textit{Gbps})}                                                                                                                                                                          & \multicolumn{1}{c|}{\multirow{3}{*}{\textbf{\% Sav.}}} \\ \cline{2-7}
\multicolumn{1}{|c|}{}                                                                               & \multicolumn{2}{c|}{\textbf{Code Rate$ = 0.9$}}                                    & \multicolumn{2}{c|}{\textbf{Code Rate$ = 0.7$}}                                    & \multicolumn{2}{c|}{\textbf{Code Rate$ = 0.5$}}                                    & \multicolumn{1}{c|}{}                                  \\ \cline{2-7}
\multicolumn{1}{|c|}{}                                                                               & \multicolumn{1}{c|}{\textbf{w/o cach.}} & \multicolumn{1}{c|}{\textbf{w/ cach.}} & \multicolumn{1}{c|}{\textbf{w/o cach.}} & \multicolumn{1}{c|}{\textbf{w/ cach.}} & \multicolumn{1}{c|}{\textbf{w/o cach.}} & \multicolumn{1}{c|}{\textbf{w/ cach.}} & \multicolumn{1}{c|}{}                                  \\ \hline
\multicolumn{1}{|c|}{$0.01$}& \multicolumn{1}{|c|}{$0.074$}  & \multicolumn{1}{|c|}{$0.018$} & \multicolumn{1}{|c|}{$0.095$} & \multicolumn{1}{|c|}{$0.023$} & \multicolumn{1}{|c|}{$0.133$} & \multicolumn{1}{|c|}{$0.033$} & \multicolumn{1}{|c|}{$75.00$}  \\ \hline
\multicolumn{1}{|c|}{$0.1$} & \multicolumn{1}{|c|}{$0.240$}  & \multicolumn{1}{|c|}{$0.185$} & \multicolumn{1}{|c|}{$0.309$} & \multicolumn{1}{|c|}{$0.238$} & \multicolumn{1}{|c|}{$0.433$} & \multicolumn{1}{|c|}{$0.333$} & \multicolumn{1}{|c|}{$23.07$}  \\ \hline
\multicolumn{1}{|c|}{$0.2$} & \multicolumn{1}{|c|}{$0.425$}  & \multicolumn{1}{|c|}{$0.370$} & \multicolumn{1}{|c|}{$0.547$} & \multicolumn{1}{|c|}{$0.476$} & \multicolumn{1}{|c|}{$0.766$} & \multicolumn{1}{|c|}{$0.666$} & \multicolumn{1}{|c|}{$13.04$}  \\ \hline
 \multicolumn{1}{|c|}{$1$}   & \multicolumn{1}{|c|}{$1.907$}  & \multicolumn{1}{|c|}{$1.851$} & \multicolumn{1}{|c|}{$2.452$} & \multicolumn{1}{|c|}{$2.380$} &\multicolumn{1}{|c|}{$3.433$} & \multicolumn{1}{|c|}{$3.333$} & \multicolumn{1}{|c|}{$2.91 $}  \\ \hline	\end{tabular}
\end{table*}

Tables~\ref{table:tput:lte} and~\ref{table:tput:docsis} compare the
downstream fronthaul transmission bitrate requirements for I/Q transmissions in
an FFT-split system without and with caching of the repetitive I/Q QAM
symbols for different packet traffic payloads (intensities) $\rho_L$
and $\rho_C$ and code rates of $0.9$, $0.7$, and $0.5$, for the LTE
and DOCSIS systems, respectively.  Tables~\ref{table:tput:lte}
and~\ref{table:tput:docsis} also report the corresponding transmission
bitrate reductions (in percent) achieved by caching the repetitive I/Q
QAM symbols.  Based on the evaluations in Sec.~\ref{sec:caching}, we
consider an I/Q QAM symbol overhead of~$7$~\% in LTE, including RS
tones, PBCH, PSS/SSS, and SIB, for a system with a bandwidth somewhat
below 20~MHz.  For DOCSIS we consider a $3$~\% overhead due to
continuous and scattered pilots, approximating the 2.9~\% found in
Eqn.~(\ref{eq:cable}).  The actual payload traffic rates are based on
wireless and cable link capacities of $R_L = R_C = 1$~Gbps, e.g., for
the traffic intensity $\rho_L = 0.01$, the actual LTE payload traffic
rate is $\rho_L \times R_L = 10$~Mbps.  The fronthaul I/Q data rate
originating from the payload traffic depends on the QAM size and code
rate of the system. With $K = 10$~bits required to represent each
complex and real part of a QAM I/Q symbol, the fronthaul transmission
bitrate required for the payload data can be evaluated as
\begin{equation}
  R^F_{\rho, \textrm{Payload}} =
  \frac{ \rho \times R } {\textrm{Code Rate} \times \textrm{QAM Size}}
     \times 2 \cdot K. \label{eq:IQrate}
\end{equation}
And the excess I/Q transmission bitrate required due to the overhead
(non-payload) can be evaluated as
\begin{equation}
  R^F_{\rho, \textrm{Overhead}} = \textrm{Overhead Percentage} \times
  R^F_{\rho = 1, \textrm{Payload }}. \label{eq:overheadIQ}
\end{equation}
The total required fronthaul transmission bitrate is the sum of
bitrates arising from overhead and payload I/Q transmissions, i.e.,
\begin{equation}
\centering
R^F_{\rho, \textrm{Total}} = R^F_{\rho, \textrm{Payload}} +
 R^F_{\rho, \textrm{Overhead}}.  \label{eq:totalIQ}
\end{equation}
Note that the system bandwidth $R_F$ from Eqn.~(\ref{RFi:eqn})
provided by the employed subcarriers  must be high
enough to accommodate the fronthaul transmission bitrate
$R^F_{\rho, \textrm{Total}}$ arising from the payload traffic
intensity $\rho$, i.e., $R^F_{\rho, \textrm{Total}} \leq R^F$.

From Table~\ref{table:tput:lte}, we observe that the reductions of the
total I/Q fronthaul data rates with caching are proportionally higher
for lower offered loads $\rho_L$.  This is because the overhead data
rate $R^F_{\rho, \textrm{Overhead}}$ is fixed at a value
corresponding to the fully loaded ($\rho_L = 1$) LTE system, whereas
the I/Q payload bitrate varies with the actual payload.  Caching
eliminates the overhead rate $R^F_{\rho, \textrm{Overhead}}$
and thus reduces the total fronthaul bitrates.  For example, for the
code rate $= 0.9$, for $\rho_L =0.01$, the total data rate without
caching is $0.296$ Gbps, which is nearly $30$ times of the offered
load $\rho_L R_L$; when $\rho_L=1$, the total fronthaul data rate without
caching is $3.962$ Gbps, which is nearly four times of the offered
load $\rho_L R_L$.  However, when caching is employed, for both loads
$\rho_L =0.01$ and $1$, the total data rates are $3.7$ and $3.33$
times of the offered load, respectively.  Higher bitrate savings can
be achieved at lower loads as compared to higher loads. For
$\rho_L=0.01$, the total savings is 87.50~\%, compared to 6.54~\%
savings for $\rho_L = 1$.

For both data rates, with and without caching, we observe linear
increases with decreasing code rates.  For example, for $\rho_L=0.01$,
the data rate without caching is increased from $0.296$ Gbps for the
code rate 0.9 to $0.380$ Gbps for the code rate $0.7$, i.e., the data
rate is increased by a factor of $0.9 / 0.7 = 1.27$.  Since both the
data rate with caching and the data rate without caching scale
linearly by a constant factor with the decreasing code rate, the
bitrate savings achieved from the overhead caching is independent of
the code rates. However, the choice of code rate for fronthaul I/Q
generation significantly affects the total data rates. Higher code
rates reduce the fronthaul requirements by lowering the total data
rate.

The throughput requirements for the DOCSIS fronthaul I/Q transmissions
presented in Table~\ref{table:tput:docsis} show similar behaviors as
the LTE results presented in Table~\ref{table:tput:lte}.  However, as
compared to the LTE fronthaul I/Q requirements for the same link
capacity of $R_C = R_L = 1$~Gbps, the DOCSIS protocol requires
relatively lower bitrates.  This is because, the DOCSIS protocol
supports a higher QAM size of 4096 ($2^{12}$) than LTE; thus DOCSIS
transports more bits per I/Q symbol transmission.  The DOCSIS overhead
percentage arising from the continuous and scattered pilot, which can
be cached at the remote node, is 3~\%. Therefore, the effective
savings in DOCSIS are relatively smaller compared to LTE.
Nevertheless, the fronthaul bitrate savings are 2.9~\% for a fully
loaded ($\rho_C = 1$) DOCSIS system and 23~\% for a 10~\% ($\rho_C =
0.1$) loaded system.

\begin{table*}[t] \centering
\caption{Total downstream LTE $+$ cable fronthaul bitrates for different
  splits: PHY (entire PHY processing at remote node), R-FFT (proposed,
  with and without caching for coding ratio 0.9), baseband
  (conventional CRAN), and passband split for range of LTE and DOCSIS
  payload traffic intensity levels $\rho_L$ and $\rho_C$ for LTE and
  DOCSIS capacities $R_L = R_C = 1$~\textrm{\normalfont Gbps}.}  \label{table:tput:overall}
\begin{tabular}{|c|c|c|c|c|c|c|c|c|c|c|c|c|c|}
\hline
\multicolumn{14}{|c|}{\textbf{Fronthaul Traffic} (\textit{Gbps})} \\ \hline
\multicolumn{2}{|c|}{\multirow{2}{*}{\begin{tabular}[c]{@{}c@{}}\textbf{PHY Split}, \textbf{payload intensity} ($\rho$), \\ payload bitrates $\rho_L R_L$ and
      $\rho_C R_C$
\end{tabular}}} & \multicolumn{6}{c|}{\textbf{FFT split $R^F_{\rho}$, Eqns.~(\ref{eq:IQrate}) and~(\ref{eq:totalIQ})}}                                                  & \multicolumn{3}{c|}{\multirow{2}{*}{\begin{tabular}[c]{@{}c@{}}\textbf{Baseband split $R^B$,} \\ \textbf{Eqn.~(2)}\end{tabular}}} & \multicolumn{3}{c|}{\multirow{2}{*}{\begin{tabular}[c]{@{}c@{}}\textbf{Passband split $R^P$,} \\ \textbf{Eqn.~(1)}\end{tabular}}} \\ \cline{3-8}
\multicolumn{2}{|c|}{} & \multicolumn{3}{c|}{\textbf{w/ caching (CR$=0.9$)}} & \multicolumn{3}{c|}{\textbf{w/o caching (CR$=0.9$)}} & \multicolumn{3}{c|}{}                                                                                         & \multicolumn{3}{c|}{}                                                                                         \\ \hline
\textbf{LTE} ($\rho_L$)& \textbf{DOCSIS} ($\rho_C$)&\textbf{LTE}& \textbf{DOC.}& \textbf{Total}&\textbf{LTE}& \textbf{DOC.}&\textbf{Total}&\textbf{LTE}&\textbf{DOC.}& \textbf{Total}&\textbf{LTE}&\textbf{DOC.}& \textbf{Total}\\ \hline
$0.01$         & $0.01$          & $0.037$  & $0.018$  & $0.055$  & $0.296$  & $0.074$  & $0.370$   & \multirow{4}{*}{$18.45$} & \multirow{4}{*}{$8.192$} & \multirow{4}{*}{$26.642$} & \multirow{4}{*}{$40$} & \multirow{4}{*}{$20$} & \multirow{4}{*}{$60$} \\ \cline{1-8}
$0.10$         & $0.10$          & $0.370$  & $0.185$  & $0.555$  & $0.629$  & $0.240$  & $0.869$   &                       &                        &                        &                     &                     &                     \\ \cline{1-8}
$0.20$         & $0.20$          & $0.740$  & $0.370$  & $1.110$  & $1.000$  & $0.425$  & $1.425$   &                       &                        &                        &                     &                     &                     \\ \cline{1-8}
$1.00$         & $1.00$          & $3.333$  & $1.851$  & $5.184$  & $3.926$  & $1.907$  & $5.833$   &                       &                        &                        &                     &                     &                     \\ \hline \end{tabular}
\end{table*}
\subsection{Total LTE $+$ Cable Fronthaul Bitrate for Different
  Function Splits} \label{eval_tot_br:sec}
The downstream fronthaul transmission
bitrate requirements to concurrently support LTE and DOCSIS
deployments over a shared optical infrastructure are shown in
Table~\ref{table:tput:overall}. The FFT split, baseband, and passband
fronthaul bitrates are evaluated based on
Eqns.~(\ref{eq:IQrate})$-$(\ref{eq:totalIQ}) and
(\ref{eq:passband})$-$(\ref{eq:freq:symb}). For the purpose of the
evaluation, we consider $W = 1$ antenna, a code rate (CR) of $0.9$,
carrier frequencies of $f_c = 2$~GHz and $1$~GHz, sampling frequencies
of $f_s = 30.72$~MHz and $204.8$~MHz, symbol durations of
$T = 66.7~\mu$s and $20~\mu$s, link capacities of $R = 1$~Gbps, and cached
overhead of $7~\%$ and $3~\%$ for LTE and DOCSIS, respectively.  We
observe from Table~\ref{table:tput:overall} that the bitrates decrease
as the position of the function split is moved from passband (i.e.,
remote DAC/ADC) to remote-PHY (i.e., from right to left in
Fig.~\ref{fig_split_phy}).

The passband bitrates $R^P_{\textrm{LTE}} = 40$~Gbps
[Eqn.~(\ref{eq:passband}) evaluated with $f_c = 2$~GHz] and
$R^P_{\textrm{DOC.}}  =20$~Gbps [Eqn.~(\ref{eq:passband}) evaluated
  with $f_c = 1$~GHz] are independent of the offered payloads $\rho_L$
and $\rho_C$.  Similarly, the baseband bitrates $R^B_{\textrm{LTE, 20
    MHz}} = 1.23$~Gbps [Eqn.~(\ref{eq:baseband}) evaluated with $f_s =
  30.72$~MHz] and $R^B_{\textrm{DOC.}} = 8.19$~Gbps
[Eqn.~(\ref{eq:baseband}) evaluated with $f_s = 204.8$~MHz] for
baseband I/Q time sample transport are independent of the offered
payloads. The LTE baseband bitrate $R^B_{\textrm{LTE, 20 MHz}}=
1.23$~Gbps is evaluated for a $20$~MHz system, which can typically
support payload bitrates up to around $R_L = 70$~Mbps with a single antenna.
Therefore, to support the payload bit rate (capacity) of $R_L =
1$~Gbps, the LTE system needs to be scaled up by a factor of at least
$15$, e.g., to an LTE system with 2 antennas, 256 QAM, and $100$~MHz
bandwidth, which can support the 1~Gbps
bitrate~\cite{mansour2015optimized}. Thus, the effective LTE baseband
bitrate to support $1$~Gbps payload is $15 \cdot R^B_{\textrm{LTE, 20
    MHz}} = 15 \cdot 1.23 = 18.45$~Gbps.  The FFT split fronthaul
bitrates with and without caching are derived from
Tables~\ref{table:tput:lte} and~\ref{table:tput:docsis}.
We observe from Table~\ref{table:tput:overall} that for a system
(without caching) loaded at 10~\% ($\rho_L = \rho_C = 0.1$), the R-FFT
approach reduces the fronthaul bitrate to 0.87~Gbps compared to
26.64~Gbps with the conventional baseband split; thus, the R-FFT
approach reduces the fronthaul bitrate to one thirtieth compared to
the conventional CRAN baseband split for this lightly loaded scenario.
For a fully loaded ($\rho_L = \rho_C = 1$) system, R-FFT reduces the
fronthaul bitrate to roughly one fifth of the baseband split.

\subsection{Delay Evaluation}
\begin{figure*}[t] \centering \begin{tabular}{cc}
\includegraphics[width=3.1in]{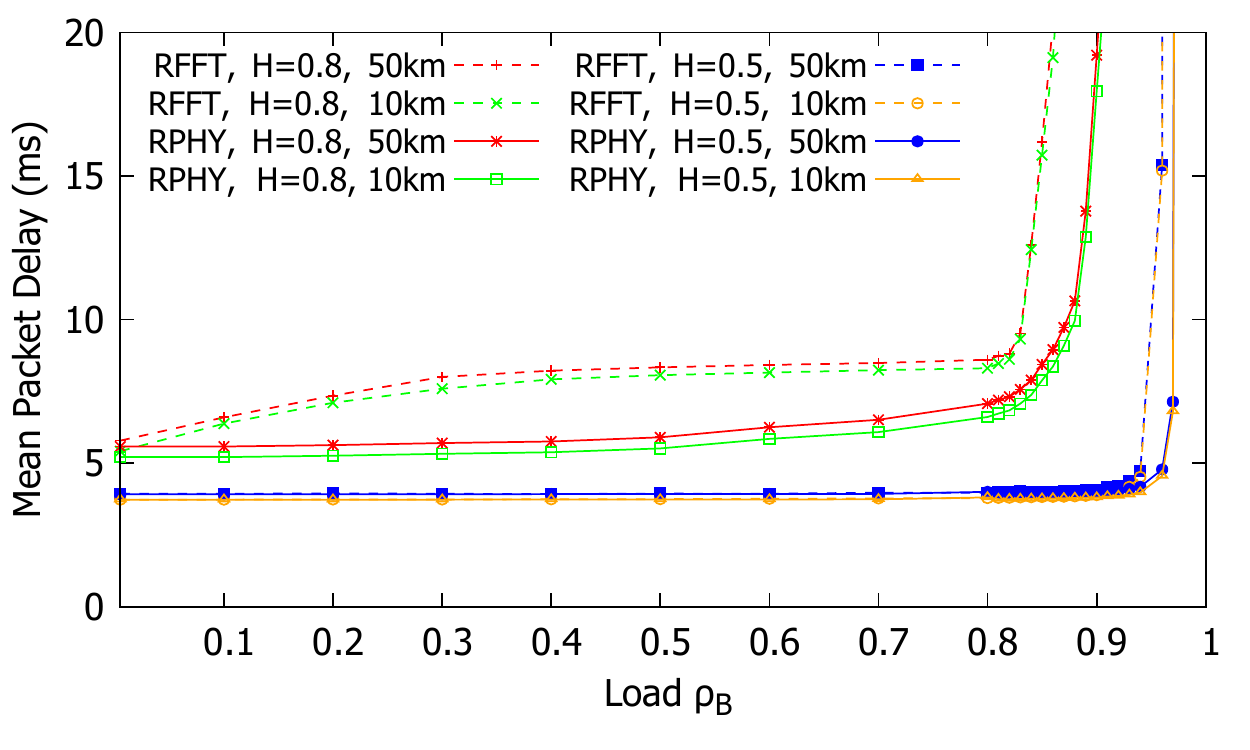} \vspace{-0.1cm}
&\includegraphics[width=3.1in]{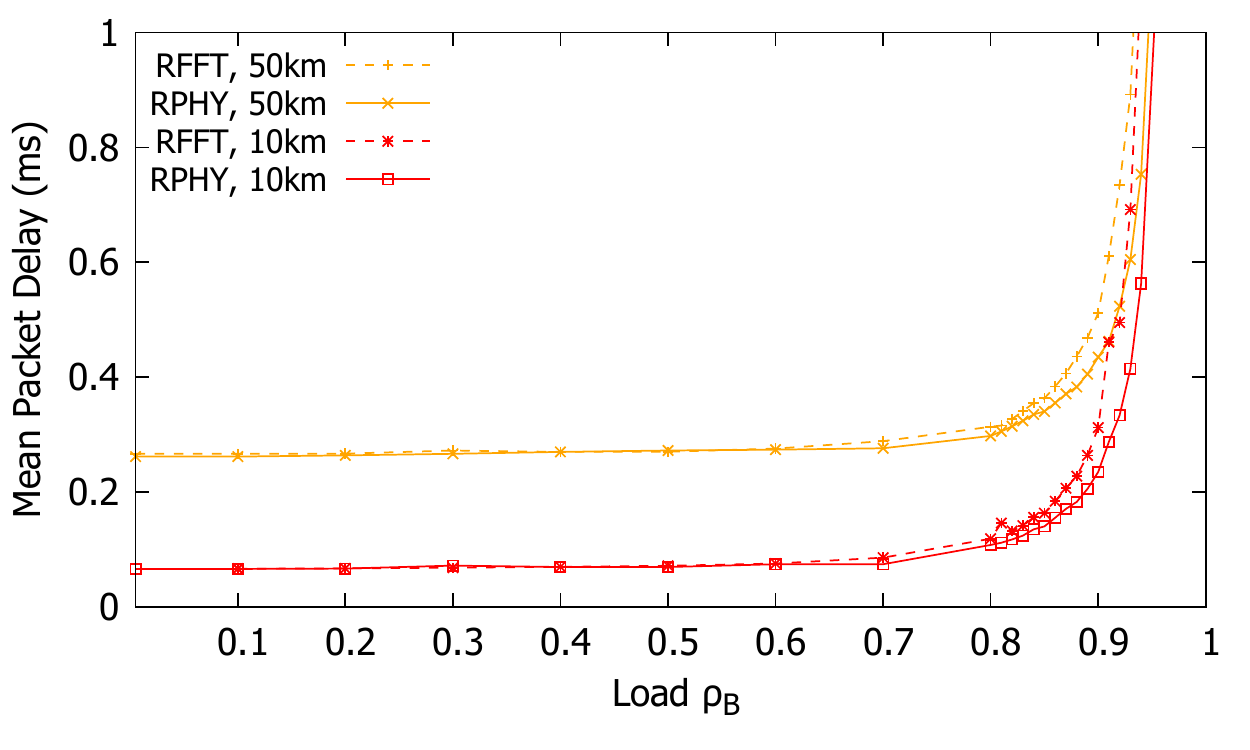} \vspace{-0.1cm}\\
		\vspace{+0cm}
\scriptsize (a) DOCSIS Delay (CMs to headend), $\rho_C = 0.2$
&\scriptsize (b) LTE Delay (remote node to BBU), $\rho_C = 0.2$, $H=0.5$\\
\includegraphics[width=3.1in]{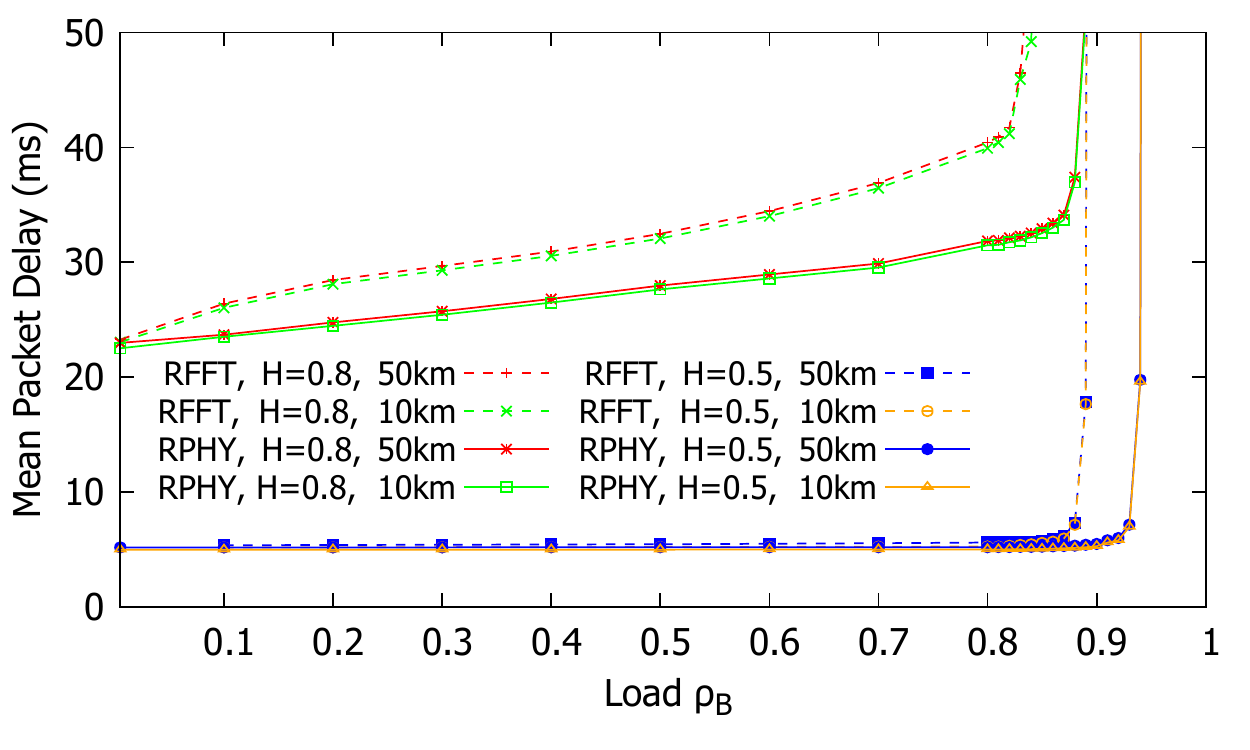} \vspace{-0.1cm}
&\includegraphics[width=3.1in]{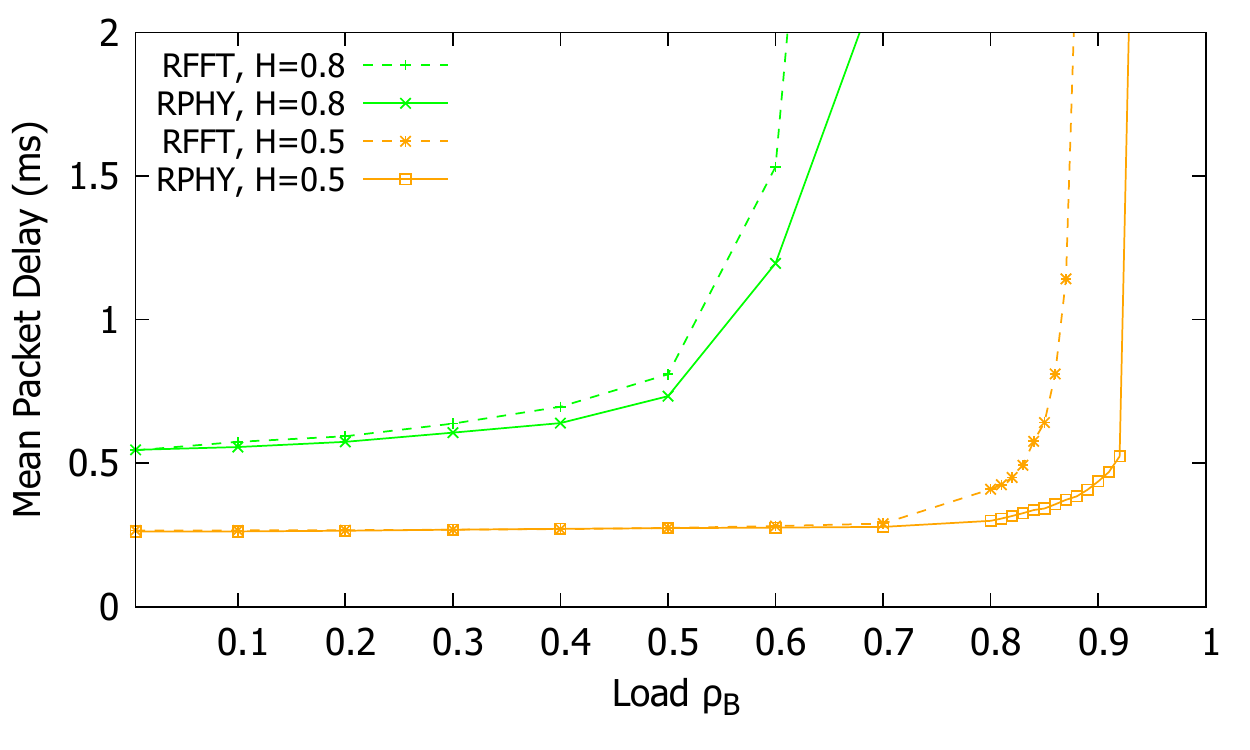} \vspace{-0.1cm}\\
\scriptsize (c) DOCSIS Delay (CMs to headend), $\rho_C = 0.6$
&\scriptsize (d) LTE Delay (remote node to BBU), $\rho_C=0.6$, $d=50$~km\\
	\end{tabular}
\caption{Mean upstream DOCSIS and LTE packet delays for R-FFT and
  R-PHY cable system supporting a prescribed load $\rho_B$ of baseband
  LTE traffic.}  \label{fig_rfft_perf}  \vspace{-0.3cm}
\end{figure*}
We proceed to illustrate the delay implications of the proposed R-FFT
deployment in comparison to the existing R-PHY deployment.  In
particular, we consider transitioning a DOCSIS cable system from R-PHY
to R-FFT operation, while sharing the fronthaul link with a fixed LTE
CRAN deployment that transmits a prescribed traffic load (intensity)
$\rho_B$ (relative to the fronthaul transmission bitrate $R_o$) of
baseband time domain I/Q sample data.

\subsubsection{Simulation Set-up}
We developed a simulation framework in the discrete event simulator
OMNET++ to model the DCCAP cable architecture of the HFC network. A
remote cable node, i.e., R-PHY or R-FFT node, is connected to the
headend through an optical fiber with CIN distance $d$ and
transmission bitrate $R_o=10$~Gbps.  We vary the CIN distance $d$
between 10 and 50~km to cover the distances of typical real deployment
scenarios.

200~cable modems (CMs) are connected to the remote node through an
analog broadcast cable.  The distances from CMs to the remote node are
uniformly distributed between 1 and 2~km in our simulations, and the
CMs are polled in shortest propagation delay order~\cite{mcg2010sho}.
Each CM has an infinite buffer in the simulation model and
independently generates self-similar traffic with varying levels of
burstiness characterized by the Hurst parameter $H$ with an average
packet size of $472$~byte.  The Hurst parameter $H=0.5$ corresponds to
Poisson traffic, and the burstiness increases for increasing $H$.  We
consider $H = 0.8$ as typical Hurst parameter for self-similar traffic
in our simulations.  The DOCSIS~3.1 protocol coordinates the cable
transmissions in the broadcast cable with the transmission bitrate
$R_C=1$~Gbps in each direction.  Throughout, we assume that 20~\% of
the cable transmission bit rate $R_C$ is occupied with contention and
maintenance slots. Thus, only 80~\% of the cable transmission bit rate
$R_C$ are available for data transmissions.  The Double Phase Polling
(DPP) protocol~\cite{cho2009dou,mer2013off,mer2016ups} controls the upstream
transmissions of the 200 distributed CMs over the shared broadcast
cable. For R-PHY, DOCSIS PHY frames are digitized and transported over
the Upstream External PHY Interface (UEPI) with prioritized CIN
transmission of the upstream transmission requests.  For R-FFT
operation, the upstream cable data is converted to frequency I/Q
symbols and transported in generic UDP packets.  An FFT size of 4096,
which corresponds to $T_C = 40~\mu$s, and QAM size of 12~bits with
code rate $0.9$ are used for converting the upstream data to frequency
domain I/Q symbols.  Each complex number representing an I/Q symbol is
digitized with $2 \cdot K = 20$~bits.

We consider the deployment of an LTE RRU at the remote cable node
(R-FFT or R-PHY).  The LTE RRU implements the conventional LTE CRAN
baseband function split, i.e., injects the baseband time domain I/Q
samples with bitrate $\rho_B R_o$ into the cable remote node.  The LTE
upstream traffic and the cable upstream traffic share the optical
transmission bitrate $R_o$ from the remote note to the headend, where
the BBU CRAN and the cable headend are implemented.  We model a
typical FIFO queue at the remote node to forward the LTE packets to
the CRAN BBU.

The average mean packet delays were sampled from over 600~s of
simulated network operation, with an additional 10~s of warm-up before
collecting samples, for each given simulation scenario.  Thus, over
180 Million packets were sampled for each considered simulation
scenario. We verified that the 98~\% confidence intervals resulting
from 10 Million simulated Poisson traffic packets were well below 2~\%
of the corresponding sample means.  We do not plot the confidence
intervals as they would not be visible.  The over 180 Million
simulated packets for each scenario result in consistent reliable mean
packet delay estimates, as demonstrated by the smooth curves in
Fig.~\ref{fig_rfft_perf}.

\subsubsection{DOCSIS Delay}
Figure~\ref{fig_rfft_perf} compares the mean upstream DOCSIS and LTE
packet delays when the cable remote node is operated as either R-FFT
or R-PHY node.  Figs.~\ref{fig_rfft_perf}(a) and (c) show the mean
cable (DOCSIS) upstream packet delay from the CMs to the headend as a
function of LTE fronthaul traffic intensity $\rho_B$, which
corresponds to the LTE I/Q sample bitrate $R^B = \rho_B R_o$ for
different optical distances $d$ and traffic burstiness levels $H$.
The cable traffic intensity is fixed at $\rho_C = 0.2$, which
corresponds to the cable traffic rate $\rho_C R_C = 0.2 \times 1$~Gbps
$=200$~Mbps.

From Figs~\ref{fig_rfft_perf}(a) and (c) we observe that the
transition from operating the cable remote node as R-PHY node to R-FFT
node slightly increases the mean DOCSIS packet delays for the bursty
$H = 0.8$ traffic, whereas the mean DOCSIS packet delays are not
visibly increased for Poisson LTE traffic loads below $\rho_B = 0.88$.
However, for very high $\rho_B$ loads, the R-FFT DOCSIS delays shoot
up to very high values at lower $\rho_B$ Poisson loads than the R-PHY
DOCSIS delays.  The underlying cause for these observations is the
increase of the cable bitrate due to the processes of I/Q conversion
and digitization. For the $9/10$ code rate, $12$~bits QAM size, and $2
\cdot K$ bits for representing the real and imaginary parts of the I/Q
samples, the cable bitrate is increased by a factor of $(10/9) \times
(1/12) \times 2 \cdot 10=1.85$ [see Eqn.~(\ref{eq:IQrate})].
There is some overhead in the uplink, e.g., for uplink pilot tones;
however, there is no overhead due to broadcast of PHY layer attributes, such as
MIB, SIB and PSS/SSS, in the uplink. We neglect therefore the uplink overhead,
which is low compared to the 1.85 fold bitrate increase due to the
I/Q conversion and digitization, in the uplink delay evaluation.

This 1.85 fold increase of the cable traffic portion on the fronthaul link
results in negligible mean delay increases for low to moderate Poisson
traffic loads.  However, for high Poisson traffic loads, the increased
cable traffic portion reduces the LTE bitrate $\rho_B$ up to which low
DOCSIS delays are achieved.  In particular, for $\rho_C = 0.6$
considered in Fig.~\ref{fig_rfft_perf}(c), the cable bitrate is
increased from $\rho_C R_C = 600$~Mbps for R-PHY to $1.85 \cdot
600$~Mbps $\approx 1.1$~Gbps; accordingly, the tolerable LTE traffic
load is reduced from close to $R_o - 600$~Mbps = 9.6 Gbps, i.e.,
$\rho_B = 0.96$, for cable R-PHY operation to only close to $\rho_B =
0.89$ for cable R-FFT operation.  Similarly, for bursty self-similar
traffic with $H=0.8$, the increase of the cable traffic portion with
R-FFT leads to more frequent temporary spikes of the total LTE plus
cable bitrate above the $R_o$ fronthaul link capacity, increasing the
mean DOCSIS packet delay compared to cable R-PHY operation.

\subsubsection{LTE Delay}
Fig.~\ref{fig_rfft_perf}(b) shows the mean LTE fronthaul packet delay
for R-FFT and R-PHY operation of the cable remote node for different
optical fronthaul distances of $d = 10$ and $50$~km.  We observe from
Fig.~\ref{fig_rfft_perf}(b) that the longer 50~km fronthaul distance
increases the LTE packet delay compared to the 10~km distance due to
the propagation delay increase [of $40 \mbox{ km}/(2 \cdot 10^8 \mbox{
    m/s})$] on the optical fiber.  We also observe that the R-FFT
cable node operation supports very slightly lower LTE traffic loads
$\rho_B$ due to the increase of the cable traffic portion from the I/Q
conversion and digitization.  Fig.~\ref{fig_rfft_perf}(d) shows the
mean LTE packet delay as a function of the LTE fronthaul bitrate
$\rho_B$ for Poisson ($H = 0.5$) and bursty ($H = 0.8$) traffic.  We
observe that the bursty traffic results generally in higher LTE mean
packet delays and gives rise to pronounced delay increases for LTE
traffic loads $\rho_B$ exceeding 0.5.

Overall, the evaluations in Figs.~\ref{fig_rfft_perf}(b) and (d)
indicate that for low to moderately hight traffic loads, the LTE
traffic suffers less than 1~ms delay.  We note that according to the
LTE protocol specifications, the LTE protocol operation is tightly
coupled to a synchronous timeline.  In particular, the LTE protocol
operates based on 1~ms sub-frames.  The end-to-end network delay along
with the processing delay is accommodated by a 4~ms separation between
an uplink request and the corresponding downlink transmission.  Thus,
in the R-FFT implementation, the total delay (network and processing),
including the jitter variations of I/Q data between BBU and the R-FFT
node, can typically be readily accommodated within the uplink-downlink
time separation on the operational timeline of the LTE protocol.

We note that the delay evaluations in this section considered the
transition of the cable remote node from R-PHY to R-FFT operation
while keeping the LTE CRAN operation unchanged.  In particular, the
cable traffic bitrate increased from the PHY payload $\rho_C R_C$ to
the FFT split bitrate [which corresponds to $R^F_{\rho,
    \rm{Payload}}$, Eqn.~(\ref{eq:IQrate})], while the LTE traffic
bitrate stayed unchanged at the baseband split rate $R^B$
[Eqn.~(\ref{RBi:eqn})].  The presented delay results represent
therefore a conservative assessment of the proposed R-FFT operation in
that a consequent transition to R-FFT operation that includes the
transition from the conventional CRAN baseband split to the proposed
R-FFT split would reduce the LTE traffic portion.  That is, the LTE
traffic portion would be reduced from the baseband split bitrate $R^B$
[Eqn.~(\ref{RBi:eqn})]
to the FFT split bitrate $R^F$ [Eqns.~(\ref{eq:IQrate})--(\ref{eq:totalIQ}),
  resp.~Eqn.~(\ref{RFi:eqn})], which is a substantial bitrate reduction.
We also note that in such a consequent transition from the
conventional R-PHY operation of the cable remote node and the CRAN
(baseband split) operation of the LTE system to the proposed FFT
split, the bitrate reduction of the LTE traffic (from baseband to FFT
split) by far outweighs the cable traffic bitrate increase (from PHY
split to FFT split).  Thus, a consequent transition to the proposed
FFT split will reduce the traffic bitrates on the fronthaul link and
correspondingly reduce delays.

\section{Conclusions}
We have developed a unified cable DOCSIS and wireless cellular LTE
access network architecture with a novel Remote-FFT (R-FFT) node.  The
proposed R-FFT architecture supports both wired DOCSIS service to
cable modems and cellular wireless LTE service over the installed
hybrid fiber-broadcast cable infrastructure.  More specifically,
DOCSIS and LTE share the fronthaul fiber link from headend to R-FFT
remote node as well as the IFFT/FFT module in the R-FFT node.  The
DOCSIS cable headend and LTE baseband unit send frequency domain I/Q
symbols over the fronthaul fiber, reducing the bitrate compared to the
conventional time domain I/Q symbol transmission.  Also, the R-FFT
node caches repetitive DOCSIS and LTE QAM symbols to further reduce
the downstream bitrate requirements over the fiber link.  Whereas
conventional cloud radio access networks require the continuous
transmission of time domain I/Q symbols over the fronthaul fiber, our
R-FFT approach with caching can temporarily suspend or statistically
multiplex the downstream transmission of frequency domain I/Q symbols
if there is no downstream payload traffic.  Our evaluations indicate
that the bitrate savings achieved with QAM symbol caching increase
substantially for low payload traffic levels. For typical DOCSIS
scenarios, the caching savings increase from 2.9~\% for a full DOCSIS
load to 23~\% caching savings with a 10~\% cable traffic load. For
LTE, the savings increase from 6.5~\% for a full wireless traffic load
to 41~\% for a 10~\% LTE traffic load.

Our evaluations also indicate that for a fully loaded system without
caching, the R-FFT approach reduces the total fronthaul bitrate
required for supporting cable and LTE wireless service to roughly one
fifth of the bitrate for the conventional baseband approach of
transmitting time-domain I/Q symbols.  For 10~\% cable and LTE traffic
load levels, our R-FFT approach reduces the fronthaul bitrate in each
direction (upstream and downstream) to approximately $1/30$ of the
conventional baseband approach.  We have also demonstrated that
transitioning a conventional R-PHY cable remote node to an R-FFT
remote node (while keeping the LTE baseband operation unchanged)
incurs only minute delay increases.  The transition to cable R-FFT
allows for the flexible efficient execution of all physical layer
processing steps (except the FFT, DAC, and upconversion) in software
on generic computing hardware at the headend, reducing the cost and
complexity of the remote node.

We note that the proposed R-FFT network approach aligns closely with
the main 5G technology development trends~\cite{agiwal2016next,
  als2017evo, gos2015fix, gou20175g}.  One main trend in 5G technology
development and deployment, especially in the fronthaul, backhaul, and
core networks, is to unify the heterogeneous access networks. Our
R-FFT approach to integrate the cable and traditional cellular
networking in the access domain is consistent with the 5G principles
of unifying the heterogeneous access networks. Another important
aspect of 5G is softwarization of traditional network applications,
such as policy enforcement and virtualization of network functions,
e.g., packet gateway functions. Towards this end, the primary goals of
the CRAN and CCAP architectures are to softwarize and virtualize
functions of cellular and cable networks.  Thus, our proposed R-FFT
architecture is overall closely aligned with the main directions of 5G
technology progress and deployment.

There are several exciting directions for future research on unifying
broadcast cable and cellular wireless access.  One particularly
important direction is to investigate how Internet of Things (IoT)
applications and traffic flows, which consist typically of small
intermittently transmitted data sets, can be efficiently served.
Future research should investigate the quality of service and quality
of experience achieved over the R-FFT network for IoT applications as
well as a wide range of general applications that require access
network transport.  Additional caching mechanisms may be useful in
efficiently serving very large numbers of such intermittent IoT flows.
Another direction is to examine and improve the interactions of the
R-FFT remote nodes and headends (BBUs) with the corresponding
metropolitan area
networks~\cite{bre2017tel,mai2003hyb,scheu2003wave,yang2003gen,zhe2017met} and
radio backhaul (core)
networks~\cite{jab20165g,mun2017tra,thy2016sdn}.
Moreover, a prototype implementation of the R-FFT approach and
evaluations through measurements in the prototype R-FFT network are an
important directions for future work.

\bibliographystyle{IEEEtran}


\begin{IEEEbiography}{Akhilesh Thyagaturu} 
	is an Engineer at Intel Mobile Communications, San Diego, CA, USA,
	and an Adjunct Faculty in the School of Electrical, Computer, and
	Energy Engineering at Arizona State University (ASU), Tempe. He
	received the Ph.D. in electrical engineering from Arizona State
	University, Tempe, in 2017.
	He serves as reviewer for various journals
	including the \textit{IEEE Communications Surveys \& Tutorials},
	\textit{IEEE Transactions of Network and Service Management}, and
	\textit{Optical Fiber Technology}.  He was with Qualcomm Technologies
	Inc., San Diego, CA, USA, as an Engineer from 2013 to 2015.
\end{IEEEbiography}

\begin{IEEEbiography}{Ziyad Alharbi} 
is a researcher at King Abdulaziz City for Science and Technology
(KACST), Riyadh, Saudi Arabia. He received his B.Sc. degree in
Electrical Engineering from King Fahd University of Petroleum and
Minerals, Saudi Arabia, and his M.S. in Electrical Engineering from
Arizona State University, Tempe. Currently, he is working towards his
Ph.D. in Electrical Engineering at Arizona State University. He serves
as reviewer for various journals including \textit{IEEE Communications
Surveys \& Tutorials}, \textit{Computer Networks}, and
\textit{Optical Switching and Networking}
\end{IEEEbiography}

\begin{IEEEbiography}{Martin Reisslein} (A'96-S'97-M'98-SM'03-F'14)
is a Professor in the School of Electrical, Computer, and Energy
Engineering at Arizona State University (ASU), Tempe. He received the
Ph.D. in systems engineering from the University of Pennsylvania in
1998. He currently serves as Associate Editor for the \textit{IEEE
  Transactions on Mobile Computing}, the \textit{IEEE Transactions on
  Education}, and \textit{IEEE Access} as well as \textit{Computer
  Networks} and \textit{Optical Switching and Networking}. He is
Associate Editor-in-Chief for the \textit{IEEE Communications Surveys
  \& Tutorials} and chairs the steering committee of the \textit{IEEE
  Transactions on Multimedia}.
\end{IEEEbiography}

\end{document}